\documentclass[preprint2]{aastex}
\usepackage{natbib}
\usepackage{graphicx}
\usepackage{url}
\usepackage{amsbsy}		

\usepackage{lineno}
\usepackage{t1enc}			

\renewcommand{\Jupiter}{{\rm J}}
\renewcommand{\Earth}{\oplus}
\newcommand{\p}{\phantom{0}}

\newcommand{\muas}{\,$\mu$as}

\shorttitle{Astrometric exoplanet detection with Gaia}
\shortauthors{Perryman et al.}

\begin{document}

\title{Astrometric exoplanet detection with Gaia}

\author{Michael Perryman\altaffilmark{1,2}, Joel Hartman, G{\'a}sp{\'a}r {\'A}.~Bakos\altaffilmark{3,4}}
\affil{Department of Astrophysical Sciences, Peyton Hall, Princeton, NJ~08544}

\and 

\author{Lennart Lindegren}
\affil{Lund Observatory, Lund, Box~43, 22100 Sweden}

\altaffiltext{1}{Bohdan Paczy{\' n}ski Visiting Fellow}
\altaffiltext{2}{Adjunct Professor, School of Physics, University College Dublin, Ireland}
\altaffiltext{3}{Alfred P.~Sloan Fellow}
\altaffiltext{4}{Packard Fellow}

\begin{abstract}
We provide a revised assessment of the number of exoplanets that should be discovered by Gaia astrometry, extending previous studies to a broader range of spectral types, distances, and magnitudes. Our assessment is based on a large representative sample of host stars from the TRILEGAL Galaxy population synthesis model, recent estimates of the exoplanet frequency distributions as a function of stellar type, and detailed simulation of the Gaia observations using the updated instrument performance and scanning law. We use two approaches to estimate detectable planetary systems: one based on the S/N of the astrometric signature per field crossing, easily reproducible and allowing comparisons with previous estimates, and a new and more robust metric based on orbit fitting to the simulated satellite data. 

With some plausible assumptions on planet occurrences, we find that some 21\,000 ($\pm6000$) high-mass ($\sim1-15M_\Jupiter$) long-period planets should be discovered out to distances of $\sim$500\,pc for the nominal 5-yr mission (including at least 1000--1500 around M~dwarfs out to 100\,pc), rising to some 70\,000 ($\pm20\,000$) for a 10-yr mission. We indicate some of the expected features of this exoplanet population, amongst them $\sim$25--50 intermediate-period ($P\sim2-3$\,yr) transiting systems. 
\end{abstract}


\clearpage

\section{Introduction}
\label{sec:introduction}

The current exoplanet census stands at around 1800\footnote{\footnotesize As of 2014 September~1, \url{exoplanet.eu} lists 1821 confirmed planets in 1135 systems, NASA's \url{exoplanetarchive.ipac.caltech.edu} lists 1743 (along with $\sim$4000 Kepler transit candidates), while the more restrictive \url{exoplanets.org} lists 1516 confirmed (with 1492 good orbits).}, with some 600 discovered from radial velocity measures, and most of the others from photometric transits. Only two (massive) astrometric discoveries have been claimed \citep{2010AJ....140.1657M,2014A&A...565A..20S}, while orbit constraints for previously-known systems are provided by Hipparcos \citep[e.g.][]{2011A&A...527A.140R,2011A&A...528L...8S} and HST--FGS astrometry \citep[e.g.][]{2010ApJ...715.1203M}. 

The astrometric detectability and characterization of exoplanets should change quantitatively with Gaia, which was launched on 2013 December~19 and began routine operations in 2014 August.  Previous work has estimated the potentially detectable numbers, both from periodic transit searches in its high-accuracy multi-epoch photometry, and independently from the astrometric displacement of the host star. 

Photometric transit detection, demonstrated {\it a posteriori\/} in the comparable Hipparcos photometry for HD~209458\,b \citep{2000A&A...355..295R} and HD~189733\,b \citep{2005A&A...444L..15B}, has been considered both for Hipparcos \citep{2000ApJ...532L..51C,2000ApJ...545.1064L,2002ApJ...564..495J,2002ESASP.485..159K} and Gaia \citep{2002EAS.....2..215R,2002Ap&SS.280..139H}. Despite an accuracy of $\sim$1\,mmag per transit at $G\la14-16$ \citep{2010A&A...523A..48J}, the low cadence makes the discovery of new transiting planets non-trivial. \citet{2012ApJ...753L...1D}, who took into account the scanning law, Galactic structure models, and detection limits to $\la$16\,mag, concluded that the low cadence and relatively small number of measurements gives a limit on the detectable orbital period of $P\la 10$\,d, and a resulting total number of expected discoveries from Gaia {\it photometry\/} of between one thousand and several thousand.

The expected number of {\it astrometric\/} planet detections was superficially estimated at the time of the mission acceptance in 2000 at around 30\,000 \citep{2001A&A...369..339P}, based on the limited knowledge of exoplanet occurrences then available, and on the higher astrometric accuracies (by a factor roughly~2) targeted at the time. Improved studies have since been undertaken \citep{2000MNRAS.317..211L,2001A&A...370..672Q,2001A&A...373L..21S}. The most detailed estimates have been made for subsets of the Gaia census by \citet{2008A&A...482..699C} for FGK dwarfs, and by \citet{2014MNRAS.437..497S} for M~dwarfs. 

\citet{2008A&A...482..699C} derived an estimated numerical yield for their sample based on star counts from the Besan{\c c}on Galaxy model, but constrained to $V<13$ and $d<200$\,pc to provide constant astrometic precision and hence uniform Gaia detectability thresholds for their orbit-fitting experiments. They adopted an along-scan single-epoch measurement error of $\sim$11\muas\ ($\sim8$\muas\ for successive crossings of the two fields of view), to be compared with the latest estimates of $\sim$34\muas, even for the brightest Gaia stars. They concluded that Gaia will detect $\sim$8000 giant planets ($M_{\rm p}>1-3M_\Jupiter$) around FGK stars out to semi-major axes 3--4\,AU. Their comprehensive double-blind simulations also led to a number of conclusions on exoplanet detectability and orbit reliability (also for two-planet systems). For example, they showed that planets with astrometric S/N~$>3$ per field crossing and period $P\leq5$\,yr can be detected reliably and consistently, with a very small number of false positives. At twice the detection limit, they found uncertainties in orbital parameters and masses of typically 15--20\%, while for favorable two-planet systems orbital elements will be measured to better than 10\% accuracy in some 90\% of cases, with the mutual inclination angle $\Delta i$ determined with uncertainties $\la10^\circ$. 

Restricting their considerations to M~dwarfs, \citet{2014MNRAS.437..497S} showed that Gaia should detect some 100 giant planets across the known sample of M~dwarf host stars within 30\,pc, and some 2600 detections and $\sim$500 accurate orbit determinations out to 100\,pc.

Motivated by the start of the Gaia operations, we re-assess the number of exoplanets detectable by Gaia astrometry. Our main objectives are to extend the previous studies to a wider parameter range (notably spectral type and distance), while taking account of recent estimates of exoplanet frequencies as a function of host star and planet properties. We use a comprehensive host star Galaxy population model, the latest instrument performance estimates, and detailed simulations of the satellite observations based on the scanning law. We quantify detection numbers both in terms of a simple S/N threshold per field crossing used in earlier work, as well as a more robust detection statistic based on orbit fitting. 

The paper is organized as follows. In Section~\ref{sec:principles} we summarize the essential concepts and quantities relevant to astrometric exoplanet detection with Gaia. In Section~\ref{sec:star-counts} we describe the sample of host stars used to quantify the numbers of planets detectable by Gaia astrometry drawn  from a population synthesis Galaxy star count model, and in Section~\ref{sec:occurrence} we present the assumptions on the exoplanet frequency estimates which we then use to simulate planets around each star. In Section~\ref{sec:detectability-sn} we derive preliminary detection statistics based on a simple consideration of the resulting astrometric signatures and the along-scan astrometric error appropriate for that stellar magnitude. In Section~\ref{sec:detectability-orbit} we estimate planet discovery numbers more rigorously by simulating the observations that will be made by Gaia, and quantifying exoplanet detectability based on goodness-of-fit improvements from the orbit solutions. In Section~\ref{sec:transiting} we focus on a statistically secure subset that is expected to transit, determining the distribution of transit depths, and quantifying the number of transit events that will simultaneously be present in the Gaia epoch photometry. In Section~\ref{sec:discussion} we discuss results for the subsets of FGK stars and M~dwarfs, briefly comparing our predicted yields with previous assessments, and we underline some of the uncertainties on our latest predicted numbers.

\section{Measurement and detection principles}
\label{sec:principles}

\subsection{Gaia astrometry}
\label{sec:gaia-astrom-photom}

Gaia, as for its predecessor Hipparcos, utilizes a small number of key measurement principles (observations above the atmosphere, two widely-separated viewing directions, and a uniform `revolving scanning' of the celestial sphere) to create a catalog of star positions, proper motions, and parallaxes of state-of-the-art accuracies \citep{2001A&A...369..339P}. Crucially, both missions provide {\it absolute\/} trigonometric parallaxes, rather than the relative parallaxes accessible to narrow-field astrometry from the ground. The observations are reduced to an internally consistent and extremely `rigid' catalog of positions and proper motions, but whose {\it system\/} orientation and angular rate of change are essentially arbitrary, since the measured arc lengths between objects are invariant to frame rotation. Placing both positions and proper motions on an inertial system corresponds to determining these 6~degrees of freedom (3~orientation and 3~spin components). For Gaia, they will be derived using the large numbers of observed quasars \citep{2006MNRAS.367..879C,2014ApJ...789..166P}.

On-board detection ensures that objects brighter than $G\sim20$ {\it at that measurement epoch\/} will be detected and observed astrometrically and photometrically, the latter through low-resolution spectrophotometry at the trailing edge of the astrometric field \citep[see][Figs~1--2]{2010A&A...523A..48J}. The highest photometric accuracy will come from the unfiltered $G$~band astrometric field photometry, which will range (per field crossing) from 1\,mmag or better for $G<14$\,mag to $\sim0.2$\,mag at $G=20$\,mag \citep[][Figure~19]{2010A&A...523A..48J}. 

Final astrometric accuracies (in positions, parallaxes, and annual proper motions) should be roughly constant at $\sim10$\muas\ (micro-arcsec) between $V\sim7-12$, degrading according to photon statistics to $\sim$20--25\muas\ at $V=15$, and to $\sim$300\muas\ at $V=20$ (precise values depend on photometric passband, star color, and astrometric parameter). These final accuracies result from the combination of the one-dimensional measurements throughout the mission, assembled using a global iterative adjustment \citep{2001A&A...369..339P,2011ExA....31..215O,2012A&A...538A..78L}. 

A single star at finite distance and with rectilinear space motion can be described by just 5~astrometric parameters, representing its position ($\alpha$, $\delta$), proper motion ($\mu_\alpha$, $\mu_\delta$), and parallax ($\varpi$). Any orbiting companions, including those of planetary mass, will perturb the stellar motion and result in deviations of the individual (`intermediate') astrometric data from a simple 5-parameter model. Detectability will depend on the amplitude of the deviations (Section~\ref{sec:astrometric-signature}), and the number and coverage of the individual measurements.

The number of individual field of view crossings, $N_{\rm fov}$, along with the final mission accuracies from which they are constructed, are primarily dependent on ecliptic latitude, $\beta$. This results from the satellite `scanning law', which is optimized to maintain a constant (solar) thermal payload illumination, while maximizing separability of the astrometric parameters. $N_{\rm fov}$ is independent of magnitude, and ranges between about $N_{\rm fov}\sim60$ at $\beta<10^\circ$ to about $N_{\rm fov}\sim80$ at $\beta>80^\circ$, with a maximum of about $N_{\rm fov}\sim150$ at intermediate ecliptic latitudes, $\beta\sim45^\circ$, where the scanning density is highest (Table~\ref{tab:fov-transits}). The high values of $N_{\rm fov}$ around $\beta\sim\pm45^\circ$ do not necessarily improve planet detection substantively, adding little to the number of distinct epochs and projection geometries. 

Our simulations of detection and orbit reconstruction require estimates of $\sigma_{\rm fov}$, the along-scan accuracy per field of view crossing as a function of $G$~magnitude (Table~\ref{tab:astrom-accuracy}). In terms of the centroiding accuracy for each of the 9~astrometric CCDs, $\sigma_\eta$, we adopt
\begin{equation}
\sigma_{\rm fov} = ( {\sigma_\eta^2 \over 9} + \sigma_{\rm att}^2 + \sigma_{\rm cal}^2)^{0.5} \ ,
\label{equ:sig-fov}
\end{equation}
where $\sigma_{\rm att}$ is the contribution from (both random and systematic modelling) attitude errors, and $\sigma_{\rm cal}$ is that from calibration errors. Both are assumed constant over the field crossing; we adopt $\sigma_{\rm att}=20$\muas\ \citep{2013A&A...551A..19R}, and similarly for $\sigma_{\rm cal}$ \citep{2012A&A...538A..78L}. Evidently, both are provisional pending results from the global iterative solution.

For bright stars, $G\la12$, signal saturation is avoided by CCD `gating', activated according to the star's measured brightness, allowing a reduced number of active TDI lines, and designed to result in a more-or-less constant measurement precision over the range $G\sim3-12$\,mag. 

In terms of the inverse relative number of photons in the image
\begin{equation}
z = 10^{\,0.4(\max[G,12]-15)} \ ,
\label{equ:z-g}
\end{equation}
normalized to $z=1$ at $G=15$, we use
\begin{equation}
\sigma_\eta = (53\,000\,z + 310\,z^2)^{0.5} \ ,
\label{equ:sig-eta}
\end{equation}
which is a fit to the values 92, 230, 590, 960, 1600, 2900\muas\ quoted for $G=13,\,15,\,17,\,18,\,19,\,20$ by \citet[][Table~1]{2012A&A...538A..78L}.

We complete these forms with an expression for the sky-averaged parallax accuracy, which can be approximated by
\begin{equation}
\sigma_\varpi = 1.2 \times 2.15\; \sigma_{\rm fov} / \sqrt{68.9} \,= 0.311\, \sigma_{\rm fov} \ ,
\label{equ:sig-parallax}
\end{equation}
where 2.15 is the geometric factor linking the (sky-averaged) parallax accuracy with the error per field crossing, 68.9 is the (sky-averaged) number of field crossings per star over the nominal 5-yr mission including dead time (Table~\ref{tab:fov-transits}), and 1.2 is a margin \citep{2012A&A...538A..78L}.

Values of $z$, $\sigma_\eta$, $\sigma_{\rm fov}$, and $\sigma_\varpi$, as a function of $G$~magnitude, are given in Table~\ref{tab:astrom-accuracy}.

\subsection{The astrometric signature}
\label{sec:astrometric-signature}

As a planet detection and characterization technique, astrometry aims at measuring the influence of an orbiting planet in addition to the two other classical astrometric effects: the linear path of the system's barycenter projected on the sky (the star's proper motion), and the reflex motion (the star's parallax) resulting from the Earth's orbital motion around the Sun.  Both star and planet orbit the star--planet barycenter and, after accounting for the parallax and proper motion terms, the orbit of the primary therefore appears projected on the plane of the sky as an ellipse with semi-major axis given by
\begin{equation}
a_\star=\left(\frac{M_{\rm p}}{M_\star}\right) a_{\rm p} \ ,
\label{equ:planet-mass}
\end{equation}
where $M_{\rm p}$ and $M_\star$ are the planet and star mass respectively, and $a_{\rm p}$ is the semi-major axis of the planet orbit with respect to the barycenter.

The observable for astrometric planet detection is the corresponding quantity in angular measure, generally referred to as the {\it astrometric signature}, given by
\begin{equation}
\label{equ:astrometric-signature}
\alpha = \left(\frac{M_{\rm p}}{M_\star}\right) \left(\frac{a_{\rm p}}{\rm 1~AU}\right) \left(\frac{d}{\rm 1~pc}\right)^{-1} \hspace{-5pt} {\rm arcsec} \ ,
\end{equation}
where $d$ is the distance, and $M_{\rm p}$ and $M_\star$ are in common units. The definition may also be adopted for $e\ne0$, but with detectability dependent on orbital phase (Section~\ref{sec:detectability-sn}). The effect is linearly proportional to $a_{\rm p}$ and, importantly, applies equally to hot or rapidly-rotating stars. But while the technique is most sensitive to massive planets at large~$a_{\rm p}$, measurement timescales must be proportionally long (of order of the orbital period).

\begin{figure}[ht]
\centering
\includegraphics[width=1.0\linewidth]{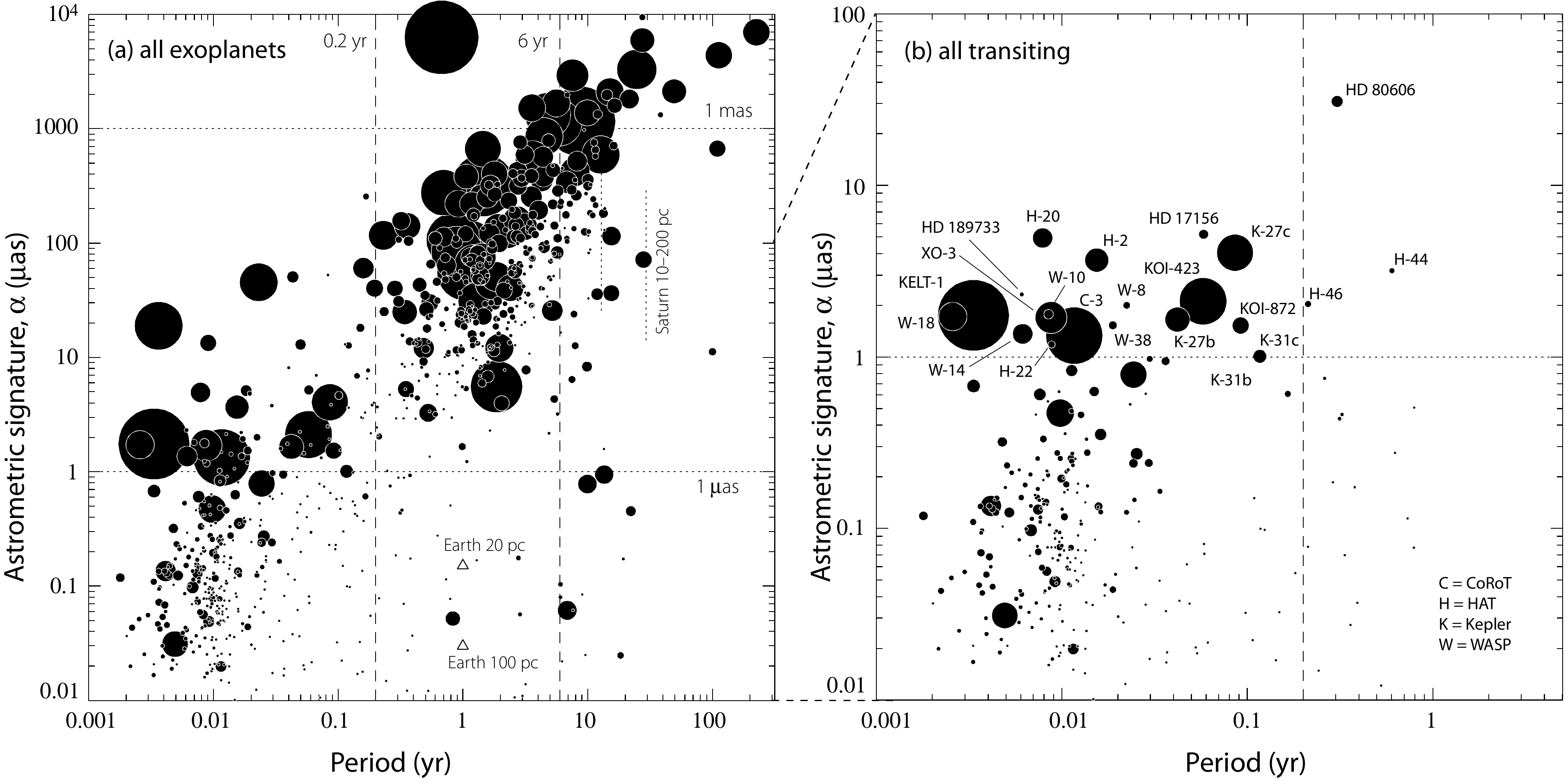}
\caption{Astrometric signature versus period calculated for the objects listed in {\tt exoplanet.eu} at 2014 September~1 for all 1821 confirmed planets (left), and for the subset of 1129 transiting planets with appropriately known data (right). Note the different scales in abscissa and ordinate. Circle sizes are proportional to planet mass; the prominent object (left) at $P=0.7$\,yr, $\alpha=6300$\muas, is the $28.5M_\Jupiter$ astrometric detection DE0823--49\,b. Unknown distances are set to $d=1000$\,pc. 
Transiting planets with $\alpha>1\,\mu$as are labelled by (abbreviated) star name, indicating the discovery instrument, both ground (H\,=\,HAT, W\,=\,WASP) and space (C\,=\,CoRoT, K\,=\,Kepler). For the transiting planets above this threshold, the unknown distance affects only Kepler--27\,b and~c, and Kepler--31\,b and~c. Assuming $d=500$\,pc, $\alpha$~would increase by a factor~2, but their astrometric motion would remain undetectable by Gaia. 
}
\label{fig:astrometric-signature-all-transits}
\end{figure}

The size of the effect calculated for all confirmed exoplanets to date (2014 September~1) is shown in Figure~\ref{fig:astrometric-signature-all-transits}a as a function of orbit period. Vertical lines illustrate the period limits between which Gaia will be most efficient in its discovery space ($0.2\la P\la6$\,yr, see Section~\ref{sec:detectability-orbit}). On the assumption that an astrometric signature of $\sim$1--3~times the parallax standard error could be detected (see Section~\ref{sec:detectability-sn}), a sizeable fraction of known systems will have their exoplanet-induced photocentric motion determined at some level by Gaia. 

Figure~\ref{fig:astrometric-signature-all-transits}b restricts the plot to the known {\it transiting\/} planets, and demonstrates that Gaia astrometry will  provide little orbital information for the majority of known transiting planets. No transiting planets have $\alpha>30$\muas, and the great majority have $\alpha\ll1$\muas. Indeed, known exoplanets with large~$\alpha$ are almost exclusively those discovered by radial velocity measurements. Even the nearest hot Jupiters will be undetected astrometrically, and the same applies to the $P\la6$\,d planets which might be discovered in the Gaia photometric data \citep{2012ApJ...753L...1D}. Gaia astrometry may nonetheless clarify the existence of massive outer companions of hot Jupiters, which have been invoked to explain their inward migration \citep[e.g.][]{2009ApJ...707..446B,2013prpl.conf2K031N,2014ApJ...785..126K}.

Various astrophysical noise sources will contribute to the accuracy of astrometric measurements in principle, but appear to lie below relevant limits in practice, and have been ignored here. These include the effects of variable stellar surface structure (star spots, plages, granulation, and non-radial oscillations) on the observed photocentre \citep[e.g.][]{2005ASPC..338...81R, 2006A&A...445..661L, 2007A&A...476.1389E, 2008NewA...13...77L, 2009ApJ...707L..73M}, relativistic modeling at $\sim1$\muas\ \citep[e.g.][]{2007A&A...462..371A}, and possible effects at optical wavelengths of interstellar and interplanetary scintillation and stochastic gravitational wave noise \citep{2001A&A...369..339P}.

\subsection{Orbit constraints from astrometric data}
\label{sec:orbit-constraints}

A 3d Keplerian orbit is described by 7~parameters, for example the classical elements $a,e,P,t_{\rm p},i,\Omega,\omega$. The semi-major axis $a$ and eccentricity $e$ specify the size and shape of the orbit. The period $P$ is related to~$a$ and the component masses through Kepler's third law, while $t_{\rm p}$~specifies the position of the object along its orbit at some reference time, generally with respect to a specified pericenter passage. The three angles 
($i$, the orbit inclination to the plane tangent to the celestial sphere; 
$\Omega$, the longitude of the ascending node; and 
$\omega$, the argument of pericenter) 
give the projection of the true orbit into the observed (apparent) orbit; they depend solely on the orientation of the observer with respect to the orbit. 

Both radial velocity and astrometry measure the host star's barycentric motion rather than that of the planet directly, and somewhat different information is provided by each measurement technique.\footnote{We recall that the sizes of the three related orbits -- the stellar orbit around the barycenter, the planet orbit around the barycenter, and the relative orbit of the planet around the star -- are in proportion $a_\star:a_{\rm p} : a_{\rm rel} = M_{\rm p} : M_\star : (M_\star+M_{\rm p})$, with $a_{\rm rel}=a_\star+a_{\rm p}$. Furthermore, $e_{\rm rel}=e_\star=e_{\rm p}$, $P_{\rm rel}=P_\star=P_{\rm p}$, the three orbits are coplanar, and the orientations of the two barycentric orbits ($\omega$) differ by $180^\circ$.
From the line-of-sight (radial) velocity variations alone, not all 7~Keplerian elements are accessible. Specifically: (i)~$\Omega$ is undetermined; (ii)~only the combination $a_\star\sin i$ is determined, with neither $a_\star$ nor $\sin i\/$ individually; (iii)~measurements provide a value for the `mass function', which for $M_{\rm p}\ll M_\star$ reduces to ${\cal M} \simeq (M_{\rm p}^3 \, \sin^3 i)/{M_\star^2}$. It follows that if $M_\star$ can be estimated from its spectral type (or otherwise), then $M_{\rm p}\sin i$ can be determined, although the planet mass remains uncertain by the unknown factor $\sin i$.}

From astrometry, all 7~Keplerian elements are accessible in some form, irrespective of the orbit inclination to the line-of-sight. Specifically, the orbit solution (including the planet location along the orbit as a function of time) gives $i$ and $\alpha$. From Equation~\ref{equ:astrometric-signature}, $a_\star$ can be determined from $\alpha$ if $d$ is known, with $a_\star+a_{\rm p}$ (and hence $a_{\rm p}$) obtained from Kepler's third law assuming that $(M_\star+M_{\rm p})\simeq M_\star$ can be estimated from the star's spectral type or from evolutionary models. Then $M_{\rm p}$ is determined from Equation~\ref{equ:planet-mass}. If the planet is invisible (the case for all but a few more massive long-period planets which have been imaged) the orbital motion of the star around the system barycenter is correctly determined by astrometry only if the star position is measured with respect to an `absolute' reference frame, which is the case for Gaia \citep{2014ApJ...789..166P}.  Astrometric measurements alone are, however, unable to identify which of the nodes is ascending, i.e.\ where the planet moves away from the observer through the reference plane, an ambiguity resolved by radial velocity observations. 

For multiple exoplanet systems, and if the orbital contributions from each can be separated, astrometry can also establish the relative inclination between pairs of orbits (e.g.\ \citealt{1981spa..book.....V}, Equation~16.5; \citealt{2008A&A...482..699C}; \citealt{2010ApJ...715.1203M}). 

Four orbit elements ($a_\star,e,\omega,t_p$) are in common between astrometric and spectroscopic orbit solutions.  Combined observations therefore further constrain and improve the 3d orbit, as well as the individual component masses \citep[e.g.][]{2009ApJS..182..205W}.

\section{Host star and planet distributions}

\subsection{Star counts}
\label{sec:star-counts}

As an input to a new set of simulations, we used the population synthesis Galaxy star count model TRILEGAL \citep[TRIdimensional modeL of thE GALaxy,][]{2005A&A...436..895G,2012rgps.book..165G}. This is based on a theoretical stellar luminosity function $\phi(M,\pmb{r},\lambda)$ [i.e., as a function of absolute magnitude $M$, Galactic position $\pmb{r}=(\ell,b,r)$, and photometric passband $\lambda$], derived from a set of evolutionary tracks, together with suitable distributions of stellar masses, ages, and metallicities. TRILEGAL (version~1.6) includes five distinct Galaxy components: the thin and thick disks, the halo, the bulge, and the disk extinction layer \citep[][Section~3.6]{2005A&A...436..895G}. The model has been calibrated with respect to a variety of observational counts, including multi-passband catalogues from very deep galaxy surveys (including CDFS, DMS, and SGP), the `intermediate-depth' near-infrared point source catalogue 2MASS, and the local stellar sample derived from Hipparcos.

A run of TRILEGAL is formally a Monte Carlo simulation in which stars are generated according to specified probability distributions. The number of stars in each bin of distance modulus is predicted according to \citep[][Equation~1]{2005A&A...436..895G}
\begin{equation}
N(m_\lambda, \ell, b) = {\rm d}m_\lambda \int_0^\infty {\rm d}r\, r^2 \,\rho(\pmb{r})\, \phi(M_\lambda,\pmb{r})\, {\rm d}\Omega \ .
\label{equ:counts}
\end{equation}
For each simulated star, the star formation rate, age--metallicity relation, and initial mass function are used to derive the stellar age, metallicity, and mass. Absolute photometry is derived via interpolation in the grids of evolutionary tracks (or isochrones), and converted to the apparent magnitudes using the appropriate values of bolometric corrections, distance modulus and extinction. All relevant stellar parameters can be retained from the simulations, including the initial and current mass, age, metallicity, surface chemical composition, surface gravity, luminosity, and effective temperature. 

The photometric system is an option in the simulation input. We selected Sloan {\it ugriz\/} combined with 2MASS {\it JHK$_{\rm s}$}. The broad-band Gaia $G$ magnitudes are then estimated as a function of Sloan $g$ and $z$ as \citep[][Table~5]{2010A&A...523A..48J} 
\begin{equation}
G = g - 0.1154 - 0.4175\,(g - z) - 0.0497\,(g - z)^2 + 0.0016\,(g - z)^3 \ ,
\end{equation}
with $\sigma=0.08$\,mag over a broad range of colors and extinction. After a number of investigation runs, the magnitude limit of our simulations was set to $r=17.5$\,mag as a compromise between retrieving a representative selection of low-luminosity stars with large astrometric signature (for reference, a magnitude limit of $r=15$ at $d=200$\,pc, corresponds to $M=0.38M_\sun$), while keeping the computational demands at a reasonable level.

We used the current version of TRILEGAL (version~1.6), with a perl script provided by L.~Girardi to automate the interaction with the www interface (\url{stev.oapd.inaf.it/trilegal}). Default model values and normalizations were used for the various Galaxy components: 
(a)~an initial mass function (IMF) given by the Chabrier log normal distribution, and a binary fraction of 0.3 with mass ratio in the range 0.7--1;
(b)~extinction according to an exponential disk of scale height 110\,pc, scale length 100\,kpc, and $A_V(\infty)=0.0378$\,mag;
(c)~solar position $R_\Sun=8.7$\,kpc, $z_\Sun=24.2$\,pc;
(d)~thin and thick disks given by $\rho\propto \exp(-R/h_R)\,f(z)$, where the vertical distribution is given by $f(z)={\rm sech}^2(0.5z/h_z)$, and where the vertical scale height of the thin disk is assumed to increase with stellar age as $h(t)=z_0(1+t/t_0)^\alpha$;
(e)~the halo is defined by an oblate $r^{1\over 4}$ spheroid, and the Galaxy is assumed to comprise a triaxial bulge;
(f)~default selections for the star-formation rate and age-metallicity relation were also used for each Galaxy component
(thus for the thin-disk: 2-step star-formation rate, with the age--metallicity relation from \citet{1998A&A...338..161F}, and $\alpha$-enhancement;
   for the thick-disk: 11--12\,Gyr constant star formation, $z=0.008$ with 0.1\,dex standard deviation, and solar-scaled abundances;
   for the halo: 12--13\,Gyr age, with [M/H] distribution from \citet{1991AJ....101.1865R};
   for the bulge: 10\,Gyr age, with [M/H] distribution from \citet{2008A&A...486..177Z} enhanced by 0.3\,dex).

Simulations were run for a field size of 1\,deg.$^2$ in $10^\circ$ steps of Galactic longitude $\ell=0-180^\circ$ (being symmetric for $180-360^\circ$) and Galactic latitude $b=-90^\circ$ to $+90^\circ$ (excluding $b=0^\circ$ where the simulations did not complete within the time imposed by the server, but including $b=\pm5^\circ$ and $b=\pm15^\circ$). For the main part of our simulations we restricted our selection to stars with $\log g>3.0$ and $\log T_{\rm eff}<4.0$, thus excluding giant and high mass stars (this approximation is discussed further in Section~\ref{sec:uncertainties}). This yielded a total of $\sim$915\,000 stars in the mass range $0.07-3.27M_\Sun$. The full-sky sample was then generated by interpolating these numbers over a mesh of 0.01\,rad in $\ell$ and $b$, finding the nearest simulated field to each grid node, and drawing the specified number of samples, with replacement, from that file. Each sample is then a simulated star.

The result is a list of $N_\star\sim260\times10^6$ stars, including binaries, whose distance distribution is shown in Table~\ref{tab:astrom-detections} ($N_\star^{\rm P14}$). Fluctuations between 300--500\,pc are due to the TRILEGAL model using discrete steps in distance modulus. Because the simulations are magnitude limited, simulated stars actually extend out to 16\,kpc. Although we consider all sample stars when simulating planets, only those within 700\,pc (for $\alpha>2\sigma_{\rm fov}$) yield detections. Of the $260\times10^6$ stars simulated, $\sim20\times10^6$ are within 700\,pc (Table~\ref{tab:astrom-detections}). 

In simulations using the Besan{\c c}on Galaxy model (version 2003, as available on the web site) to replicate the detection numbers reported by \citet{2008A&A...482..699C}, we found that the TRILEGAL star counts were some 30\% lower out to $\sim$200\,pc than those returned by the Besan{\c c}on galaxy model. A.~Robin (2014, priv.\ comm.) has attributed this to the fact that the star formation rate in the Besan{\c c}on model was assumed constant over the thin disk life time; the revised model described by \citet{2014A&A...564A.102C} is expected to correct much of this discrepancy. We have accordingly adopted the results from TRILEGAL, noting that some 30\% more exoplanets would be detected astrometrically should the star counts (and the treatment of extinction) follow more closely that predicted by the Besan{\c c}on model.

\subsection{Assumed exoplanet occurrence dependencies}
\label{sec:occurrence}

Recent developments in characterizing exoplanet occurrence frequencies as a function of planetary and stellar properties are directly relevant to our re-assessment. These include:
(a)~improved statistics of the longer-period radial velocity discoveries, allowing a more secure extrapolation for large $M_{\rm p}$ and $P$; 
(b)~larger numbers of Neptune-mass discoveries, which augments the contribution of lower-mass planets at large~$a_{\rm p}$ (while the distant limit for detection will decrease compared with those of Jupiter mass, this will be compensated by the increased numbers at smaller~$d$);
(c)~improvements in characterizing the occurrence of planets around M~dwarfs. 
At the same time we have introduced some very simple, and intentionally conservative, assumptions in cases where occurrence rates are unknown or at best poorly known empirically. 

More specifically, for each of the 260~million stars returned by the TRILEGAL simulations to $r<17.5$\,mag, we simulate the exoplanet occurrence and properties according to the following dependencies:

\noindent
(1)~binary stars: the secondary stars of binaries were ignored, and planets simulated around the primaries according to the occurrence distributions for single stars (this simplification is discussed in Section~\ref{sec:uncertainties});

\noindent
(2)~host star mass and metallicity: for giant planets around host stars with $M_\star>0.6M_\Sun$ we used the relationship between giant planet occurrence and host star mass and metallicity given by \citet{2010PASP..122..905J}. This relation was determined for stars with $0.5M_\Sun<M_\star< 2.0M_\Sun$ and [Fe/H]\,$<$\,0.4. We assume that the occurrence rate of giant planets around stars with $M_\star>2.0M_\Sun$ is equal to that at $M_\star=2.0M_\Sun$, and that the rate for stars with [Fe/H]\,$>$\,0.4 is equal to that at [Fe/H]\,=\,0.4. We extrapolate occurrence rates to stars off the main sequence (including white dwarfs), while excluding from consideration giant and high mass stars (as noted in Section~\ref{sec:star-counts}, and discussed further in Section~\ref{sec:uncertainties}). Our treatment of host star mass and metallicity for smaller planets and for planets around stars with $M<0.6M_\Sun$ is discussed below; 

\noindent
(3)~planet mass and orbital period: we assume that the planet mass and period distributions are independent of metallicity. For each star we draw planets from the joint mass--period distribution given by \citet{2008PASP..120..531C}, as determined from radial velocity surveys for GK stars. For planet masses $>0.3M_\Jupiter$ and periods $<2000$\,d, they obtained a power law fit ${\rm d}N=C\,M^\alpha \,P^\beta\, {\rm d}\ln M\,{\rm d}\ln P$ with $\alpha=-0.31\pm 0.2$, $\beta=0.26\pm0.1$, and the normalization constant $C$ such that 10.5\% of solar type stars have a planet with mass in the range $0.3-10M_\Jupiter$ and orbital period 2--2000\,d. For $M_\star>0.6M_\Sun$, we extrapolated this power law to cover the extended mass range $0.1-15M_\Jupiter$, and to orbital periods up to 10\,yr (for very wide orbits around A~stars the best constraints are currently based on direct imaging, but the semi-major axes probed lie well beyond the range of orbit period relevant for Gaia). For $P<418$\,d, the outer period to which the Kepler distributions were determined, we did not extrapolate the Doppler-based distributions below $0.3M_\Jupiter$, and instead used the Kepler results for planets of lower mass (Table~\ref{tab:planet-frequencies});

\noindent
(4)~occurrence around low-mass stars (M~dwarfs): we examined two dependencies. The first is a simplified extension of the Johnson--Cumming distributions, for which we assumed that the occurrence rate of gas giant planets around stars with $M_\star<0.5M_\Sun$ was equal to that at $M_\star=0.5M_\Sun$. The second (which we have adopted for all subsequent steps) follows the conclusions of \citet{2014ApJ...781...28M}, who found that the Cumming et al.\ mass--period power-law does not agree with the microlensing results for M~dwarfs (see also \citet{2014ApJ...791...91C} for a recent analysis reconciling the radial velocity and microlensing planet yields for M~dwarfs). From Doppler measurements of 111~M~dwarfs, they found a lower occurrence rate compared to higher mass stars, finding only $6.5\pm3.0$\% host one or more massive companions with $0<a_{\rm p}<20$\,AU and $1M_\Jupiter<M_{\rm p}<13M_\Jupiter$. Accordingly, we adopt as our baseline, and more conservative estimate for $M_\star<0.6M_\Sun$, their double power law in stellar mass and metallicity
\begin{equation}
f(M_\star, {\rm [Fe/H]}) = 0.039\, M_\star^{0.8} \, 10^{3.8\,{\rm [Fe/H]}} \ ,
\label{equ:montet1}
\end{equation}
with a dependency on $M_{\rm p}$ (and flat in $\log a_{\rm p}$), consistent with both their observations and the microlensing observations, given by
\begin{equation}
{\rm d}N   \propto \ M_{\rm p}^{-0.94}  \ {\rm d}\ln M_{\rm p}\, {\rm d}\ln a_{\rm p} \ .
\label{equ:montet2}
\end{equation}
For the same stars, $M_\star<0.6M_\Sun$, we extrapolated this power law to cover the extended mass range $0.1-15M_\Jupiter$. Using the \citet{2014ApJ...781...28M} distribution for $M_\star<0.6M_\Sun$ in place of the Johnson--Cumming distributions substantially reduces the expected Gaia planet yield for two reasons: (1)~rather than assuming that stars with $M_\star<0.5M_\Sun$ have the same planet occurrence fixed to the Johnson et al. rate at $M_\star=0.5M_\Sun$, the Montet et al.\ relation predicts that the planet occurrence rate continues to drop for lower mass stars; (2)~although the occurrence rate of small planets around M~dwarfs is higher in the Montet et al.\ distribution, the larger planets and longer period planets which Gaia would be sensitive to are less common;

\noindent
(5)~eccentricities: we assumed that eccentricities of the 3d orbits follow a Beta distribution
\begin{equation}
{\rm P}\!_{\beta}\,(e;a,b) \ \propto \ e^{a-1}\, (1-e)^{b-1} \ ,
\label{equ:kipping-beta}
\end{equation}
with $a=0.867$ and $b=3.03$, as established from fits to radial velocity observations by \citet{2013MNRAS.434L..51K}. We ignore any possible dependency on period (in practice, the planets detectable with Gaia astrometry are on wide orbits where this assumption, in the absence of tidally-circularized orbits, appears justified). The eccentricity was not used in deciding whether an astrometric signal would be detected (Section~\ref{sec:detectability-sn}), but it is used in determining the enhanced transit probability for elliptical orbits (Section~\ref{sec:transiting});

\noindent
(6)~distribution of low-mass planets: for $M_{\rm p}<0.3M_\Jupiter$, we used the joint period--radius distribution determined from the Kepler data by \citet{2013ApJ...766...81F}. This extends only out to $P=428$\,d for the largest planets, and since a power law does not appear to provide a particularly good fit, we did not extend the distributions to longer periods. Radii were converted to masses using the following empirical relation, chosen to match the \citet{2013ApJ...766...81F} occurrence rate (as a function of $R_{\rm p}$) to the \citet{2010Sci...330..653H} occurrence rate (as a function of $M_{\rm p}$) for $P<50$\,d, and with $M_{\rm p}=0.3M_\Jupiter$ at $R_{\rm p}=1R_\Jupiter$ (the same approach as used by \citet{2012ApJS..201...15H} to compare their radius distribution from Kepler to their earlier mass distribution from Doppler observations, \citealt{2010Sci...330..653H})\\[0pt]
\parbox{3cm}{
\begin{eqnarray*}
\label{equ:mass-radius}
M_{\rm p}	&=&	\phantom{0}1.08\,R_{\,\rm p}^{\,3.45}	 \\
		&=&	\phantom{0}3.17\,R_{\,\rm p}^{\,0.87} \\
		&=&			10.59\,(R_{\,\rm p}/4)^{\,2.07} \\
		&=&			24.51\,(R_{\,\rm p}/6)^{\,2.17} 
\end{eqnarray*}
}	
\hspace{0.2cm}
\parbox{5cm}{
\begin{eqnarray*}
	 &\hspace{-10pt}R_{\rm p}&\hspace{-5pt}\le 1.5 \\
1.5 < &\hspace{-10pt}R_{\rm p}&\hspace{-5pt}\le 4.0 \\
4.0 < &\hspace{-10pt}R_{\rm p}&\hspace{-5pt}\le 6.0 \\
6.0 < &\hspace{-10pt}R_{\rm p}&
\end{eqnarray*}
}
\hfill
\parbox{1cm} {\begin{eqnarray}  \label{equ:fressin-mr} \end{eqnarray}}\\
where $M_{\rm p}$ and $R_{\rm p}$ are expressed in Earth units.

\noindent
(7)~multi-planet systems: the \citet{2008PASP..120..531C} distribution uses only the most significant Doppler signal for stars with multiple planets. Carrying this over into our planet occurrence simulations, we do not account for multiple gas giant systems. The \citet{2013ApJ...766...81F} distributions based on Kepler, on the other hand, include all planets in the multi-planet systems. In this case, we have partitioned the distribution into their designated mass bins (Table~\ref{tab:planet-frequencies}), and generate at most one planet per mass bin for each star.  In practice, no planets from the parameter space covered by the \citet{2013ApJ...766...81F} distribution are recovered with $\alpha>2\sigma_{\rm fov}$, and therefore we do not recover any multi-planet systems. 

Table~\ref{tab:planet-frequencies} summarizes the fraction of stars with planets from the combination of Kepler transit and radial velocity data, as a function of mass and period. The end result is that for our 260~million simulated stars to $r=17.5$, we simulated 254~million planets (with 79~million representing additional planets within a multiple planet system), accompanying 175~million stars.

We stress that our treatment of multi-planet systems is incomplete. First, we do not allow for more than one (low-mass) planet in one mass bin. However, in practice, the entire region of parameter space covered by the \citet{2013ApJ...766...81F} distribution yields just 1~marginally detectable Gaia planet (with $M_{\rm p}=0.29M_\Jupiter$, $P=389$\,d, and $\alpha=1.15\sigma_{\rm fov}$). To this extent, our simplistic assumptions do not affect the question of astrometric detectability.  Second, we have assumed that the astrometric motion of the host star is dominated by a single massive planet. To go further is beyond the scope of this paper: current knowledge of orbit statistics for multiple gas giants at large~$a_{\rm p}$ is limited, while the astrometric motion of the star with respect to the system barycenter for multiple massive planets rapidly becomes more complex \citep[see, e.g.][Figures~2--3]{2011A&A...525A..65P}. We refer to \citet{2008A&A...482..699C} for further insight into the detectability (although not the occurrence) of multiple massive systems.

\section{Detectability based on S/N per field crossing}
\label{sec:detectability-sn}

From their numerical double-blind simulations \citet{2008A&A...482..699C} argued that a planet is detectable by Gaia if the astrometric signature exceeds some three times the accuracy of a single field crossing, $\alpha\ga 3\,\sigma_{\rm fov}$. The latter, we recall, is only a function of magnitude (it is only the scanning density that depends on ecliptic latitude). Using the relation between $\sigma_{\rm fov}$ and the sky-averaged parallax accuracy $\sigma_\varpi$ (Equation~\ref{equ:sig-parallax}) gives a rough indication that planets become detectable for $\alpha\ga \sigma_\varpi$.

While a (single) S/N threshold per field crossing provides some indication of exoplanet detection numbers, it is evidently simplistic. Shortcomings include: (i)~the number of field crossings, and their distribution in projection angle and time, is variable over the sky; (ii)~the number and distribution of geometrically-independent measurements depends on orbit period; (iii)~the detectability of elliptical orbits varies over orbit phase (cf.\ Equation~\ref{equ:astrometric-signature}); (iv)~all of these effects also depend on the actual mission duration. Deeper insight into these and other effects requires Monte Carlo-type simulations spanning a range of planetary system parameters, representative satellite observations, and the construction of some statistic quantifying detectability based on orbit modeling. We defer this more detailed treatment to Section~\ref{sec:detectability-orbit}.

We start by determining zero-order detection numbers for a range of S/N thresholds per field crossing (in which the star counts and exoplanet distributions remain fixed)
\begin{equation}
{\rm S/N} \equiv \alpha/\sigma_{\rm fov}\, >\, n\, \ ,
\label{equ:exoplanet-detection}
\end{equation}
where $n$ is a detection threshold parameter in the range $n=0.5-6$. Our justification for this approach is that candidates can be easily identified according to such a S/N criterion, yielding insights and results which can be replicated without recourse to more detailed simulations, and which can be compared directly with previous estimates.

Resulting provisional detections can then be estimated from the parameters of the 240~million simulated exoplanet systems (Section~\ref{sec:occurrence}), the resulting astrometric signature calculated on a system-by-system basis (Equation~\ref{equ:astrometric-signature}), and the simple detection criterion given by Equation~\ref{equ:exoplanet-detection}, in which the along-scan accuracy per field crossing as a function of $G$ magnitude ($\sigma_{\rm fov}$, Equation~\ref{equ:sig-fov}) is as tabulated in Table~\ref{tab:astrom-accuracy}.

Table~\ref{tab:astrom-detections} summarizes the results versus distance intervals from the Sun, as a function of S/N threshold. For comparison with previous work, $N_{\rm FGK}^{\rm C08}$ gives the number of FGK dwarfs from the Besan\c con Galaxy model as derived by \citet{2008A&A...482..699C}, and $N_{\rm det}^{\rm C08}$ gives the corresponding numbers of giant planets they detected with their criterion $\alpha>3\,\sigma_{\rm fov}$. $N_\star^{\rm P14}$~gives our star counts from the TRILEGAL model (Section~\ref{sec:star-counts}).  Subsequent pairs of $N_{\rm det}$ and $N_{\rm tran}$ give the resulting numbers of detected planets for values of the S/N threshold in the range $n=0.5-6$, and the corresponding number of predicted {\it transiting\/} astrometric detections, which we expand on further in Section~\ref{sec:transiting}.

The criterion $\alpha>3\,\sigma_{\rm fov}$ corresponds to that used by \citet{2008A&A...482..699C}, albeit with our more realistic values of $\sigma_{\rm fov}$ ($\sim$34\muas\ for $G<12$, compared to their $\sim$11\muas), while $\alpha>6\,\sigma_{\rm fov}$ corresponds to their `twice the detection limit' criterion for good orbit determination. The criterion $\alpha>1\,\sigma_{\rm fov}$ (roughly) corresponds to a $2\sigma_{\rm fov}$ detection threshold that {\it would have applied\/} to the accuracies targeted at the time of the mission acceptance by the ESA Science Programme Committee in 2000 ($\sigma_\varpi=10$\muas\ at $G=15$\,mag) compared to the current accuracies ($\sigma_\varpi=25$\muas\ at $G=15$\,mag). We note in passing that in its accepted form (before descoping), and according to this simplified detection criterion, Gaia would have detected some 20\,000 (at $3\sigma$) to 30\,000 (at $2\sigma$) planets, consistent with the preliminary estimates of around 30\,000 given at that time \citep{2001A&A...369..339P}.

\begin{figure*}[ht]
\centering
\includegraphics[width=0.9\linewidth]{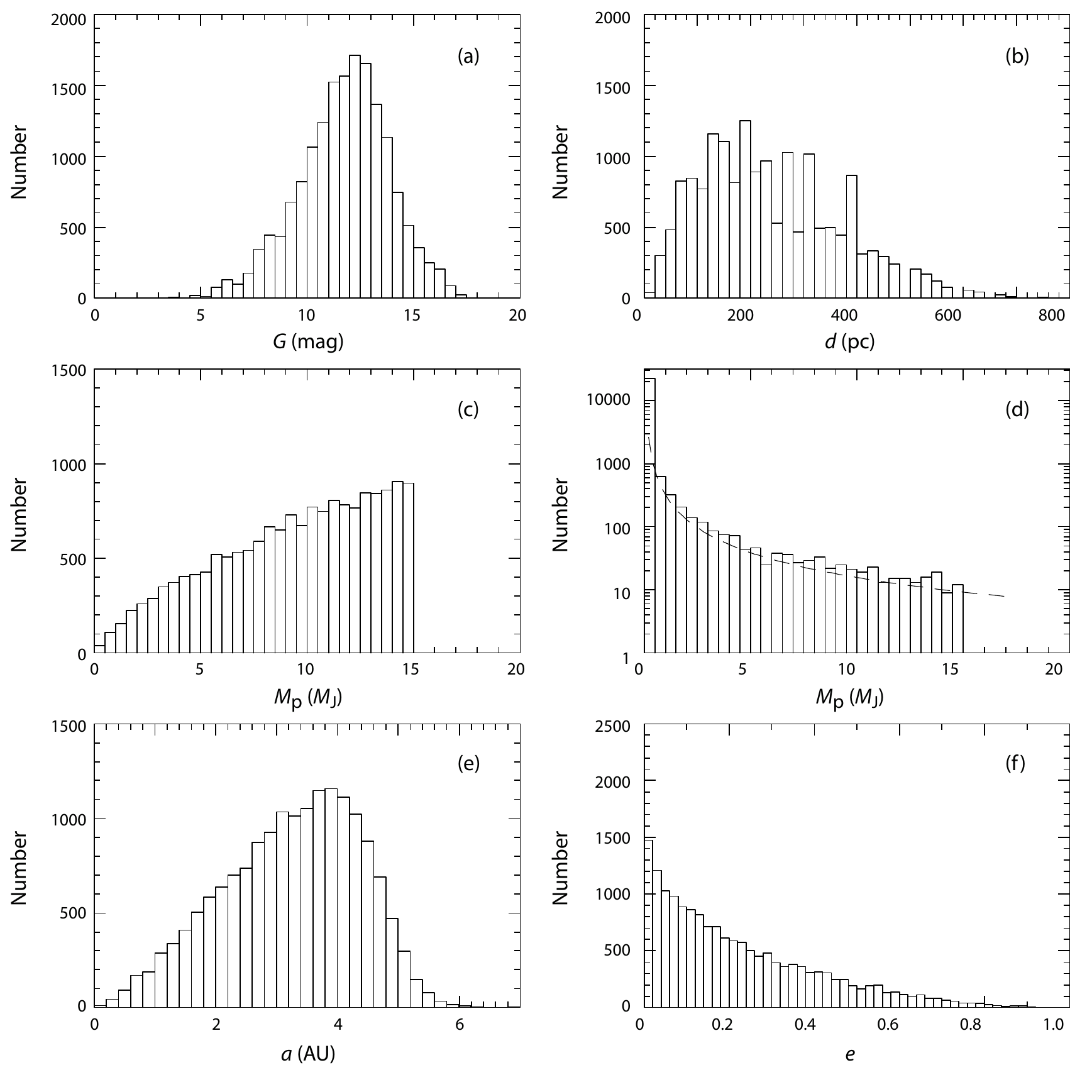}
\vspace{-20pt}
\caption{Histograms of the predicted planet detections with Gaia, for $\alpha>2\,\sigma_{\rm fov}$, as a function of:
(a)~$G$~magnitude;
(b)~distance~$d$;
and (c)~planet mass~$M_{\rm p}$, where the rising distribution up to the adopted occurrence threshold of $15M_\Jupiter$ contrasts with the mass distribution for the simulated planets (d, which is a 1 in $10^4$ re-sampling of the simulated planet distribution, showing the $M^{-1.31}_{\rm P}$ power-law behaviour as the dashed line);
(e)~semi-major axis~$a$;
(f)~eccentricity~$e$.
}
\label{fig:histogramsallplanets}
\end{figure*}

\begin{figure*}[ht]
\centering
\includegraphics[width=0.9\linewidth]{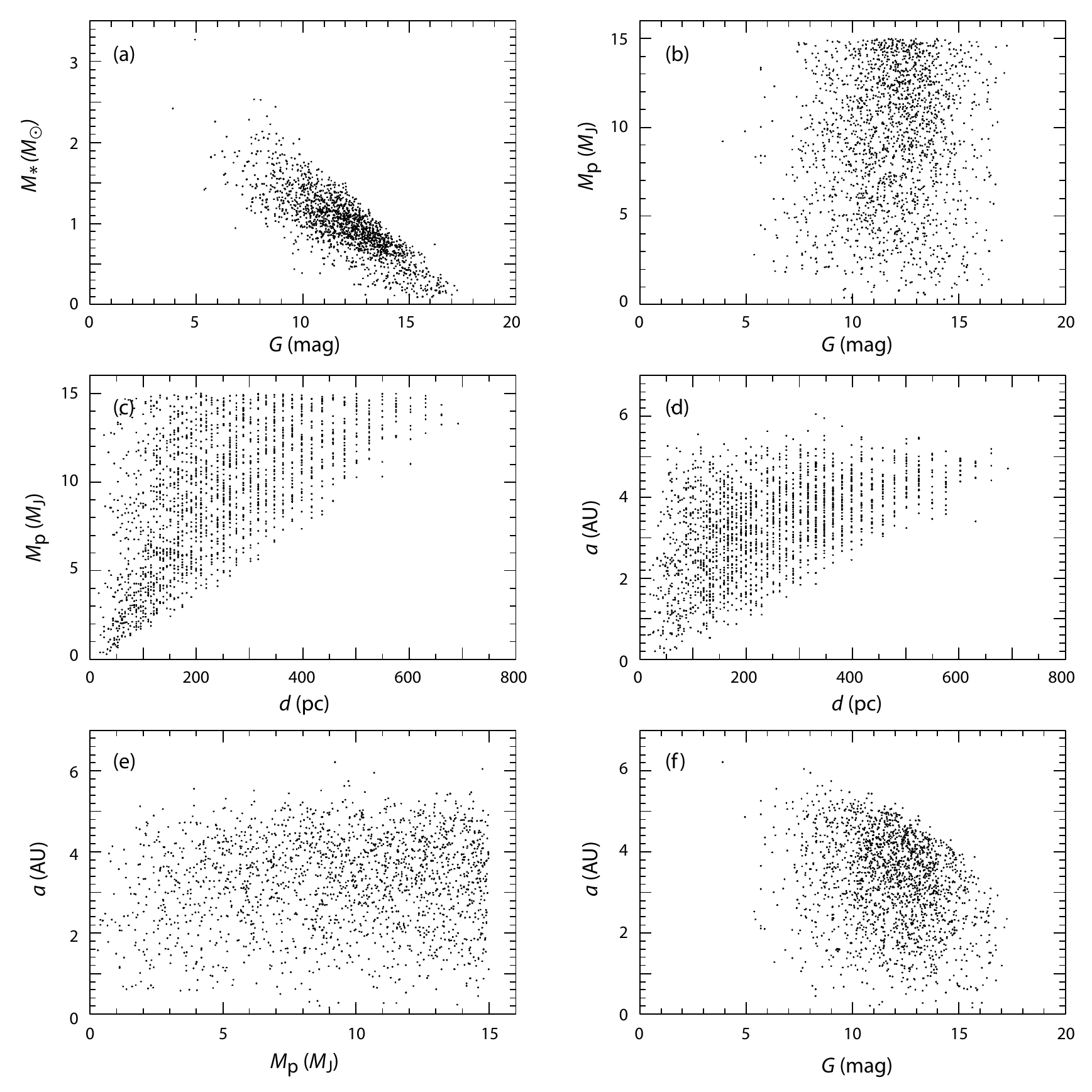}
\vspace{-10pt}
\caption{Scatter diagrams (with a random 1~in~10 sampling) of the predicted planet detections with Gaia (for $\alpha>2\,\sigma_{\rm fov}$):
(a)~fainter stars with detectable planets are of preferentially low-mass;
(b)~a broad distribution of $M_{\rm p}$ remains detectable even for the fainter stars;
(c)~the most distant planet detections are restricted to the most massive planets, with
(d)~the largest semi-major axes up to the period limit of Gaia detectability;
(e)~there is a broad range of $M_{\rm p}$ versus $a$;
and (f)~planets around the fainter stars have smaller $a$, related to the decreasing $M_\star$.
Granularity in distance is due to the discrete absolute magnitude steps in TRILEGAL.
}
\label{fig:scatterallplanets}
\end{figure*}

We will show, through the more detailed analysis in Section~\ref{sec:detectability-orbit}, that a threshold of $\alpha>2\,\sigma_{\rm fov}$ (more liberal than the $3\sigma$ condition used by \citealt{2008A&A...482..699C}), provides a reasonable approximation to the final {\it numbers\/} that we consider can be detected. Table~\ref{tab:astrom-detections} then indicates that we can expect a total of 16\,668 recovered planets around stars with $r<17.5$\,mag, $\log g>3.0$ and $\log T_{\rm eff}<4.0$. Histograms of these 16\,668 exoplanets, showing the distributions of $G$~magnitude, $d$, and $M_{\rm p}$ are given in Figure~\ref{fig:histogramsallplanets}a--c, and relevant scatter diagrams in Figure~\ref{fig:scatterallplanets}a--f. Detected planets have masses in the range $M_{\rm p}=0.12-15 M_\Jupiter$, semi-major axes in the range $a_{\rm p}=0.037-6.87$\,AU, and are around stars with masses in the range $M_\star=0.07-3.27M_\Sun$. Periods range between 7.4\,d--10\,yr, with 327 below 1\,yr, and 7390 below our provisional upper limit for orbit solutions of $\sim6$\,yr. 

The distribution of planet masses in our recovered sample (Figure~\ref{fig:histogramsallplanets}c) continues to increase to the highest planet masses included in the simulations ($15M_\Jupiter$).  This contrasts with the input power-law distribution ($\propto M_{\rm P}^{-1.31}$) of simulated planet masses (Figure~\ref{fig:histogramsallplanets}d). The increasing numbers of high-mass planets results from the competition between a falling occurrence rate for the more massive planets and the increasing astrometric signatures for larger planets, and thus a greater volume of space and number of stars for which these planets can be detected. To extrapolate much above $15M_\Jupiter$ goes beyond the scope of this study: it would require a suitable transition to the (very different) stellar binary mass ratio distribution, extending the Galactic models to include more massive objects around the faintest stars, and wider questions of the Gaia detectability of binary stars more generally.

\section{Orbit fitting and an improved detectability metric}
\label{sec:detectability-orbit}

We now turn to the problem of orbit reconstruction based on simulated Gaia data. We will show that due consideration of the orbit fit allows us to quantify detectability more rigorously. In the process we can estimate the precision that can be obtained on some of the key orbital parameters, such as the orbit inclination and period. 

\subsection{Simulated data and orbit fitting}
\label{sec:simulated-data}

We generate simulated observations of large numbers of exoplanet systems using tools provided by the Gaia AGISLab project \citep[][their Appendix~B]{2012A&A...543A..15H}. Developed as part of the astrometric global iterative solution \citep[AGIS,][]{2012A&A...538A..78L} AGISLab allows the simulation of millions of sources at the level of individual CCD transits, using a comprehensive instrument model, including the scanning law. Using this we can generate, for any target star based on its sky coordinates, a listing of all field crossings over the mission, giving the time, position angle of the scan, and the parallax factor for each observation. This, together with the assumed along-scan standard error per field crossing (as a function of~$G$), and the seven specified orbital elements of the star's reflex motion, allows us to simulate a full set of representative (`intermediate astrometry') observations.

\begin{figure}[ht]
\centering
\includegraphics[width=0.5\linewidth]{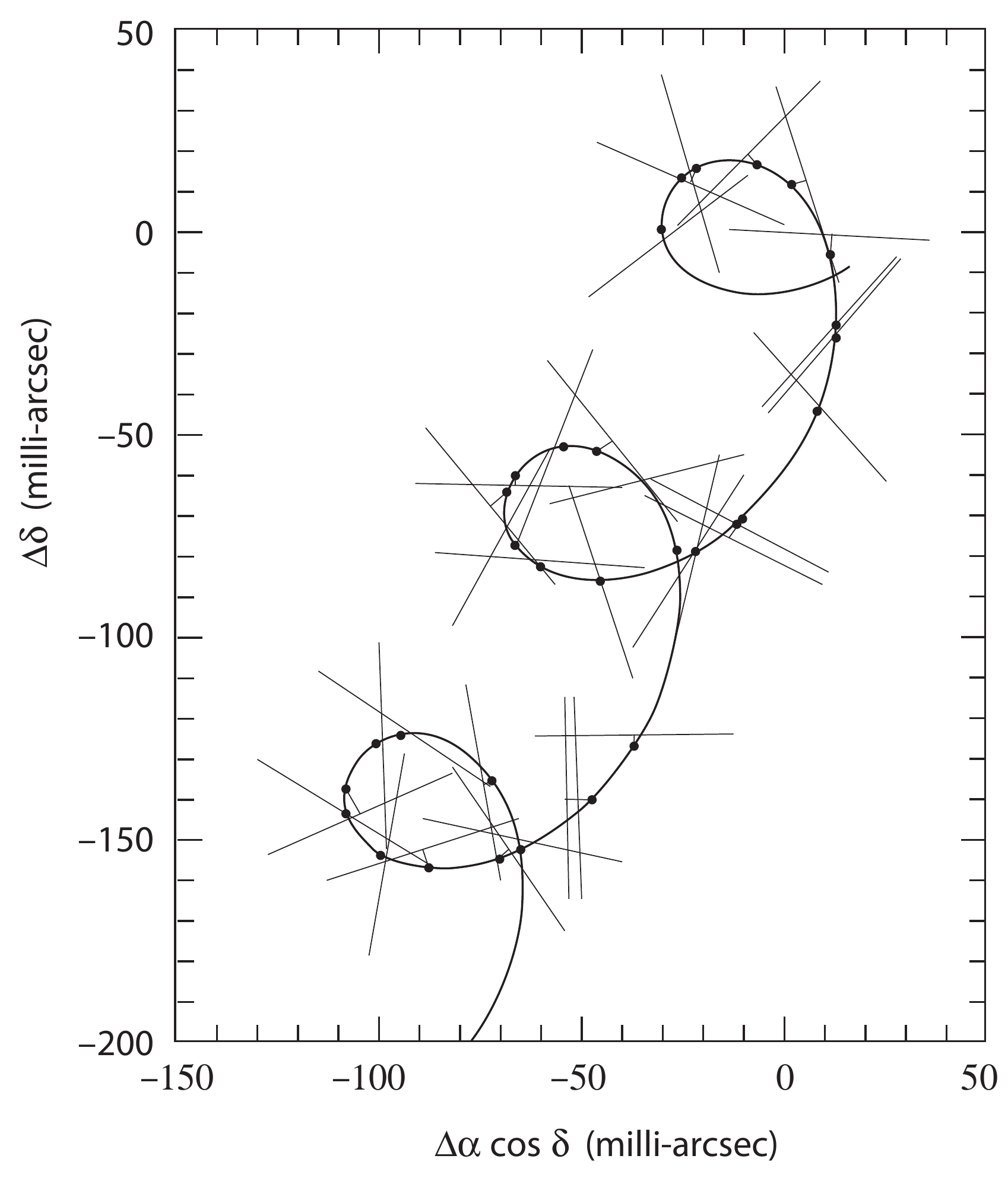}
\vspace{-5pt}
\caption{The path on the sky of a star from the Hipparcos catalogue, over 3~years, illustrating the principle of the intermediate astrometric data used for object multiplicity fitting. Each line indicates the observed position of the star at a particular epoch: because the measurement is 1-d, the precise location along each line is undetermined. Curves show the (5-parameter) modelled stellar path fitted to all measurements. The inferred position at each epoch determined by the fit is indicated by a small filled circle, and the residual by the short line joining the circle to the corresponding position line. The amplitude of the oscillatory motion gives the star's parallax, with the linear component defining the star's proper motion.
}
\label{fig:intermediate-astrometry}
\end{figure}

We then subject these simulated observations for each system to a least-squares orbit fitting algorithm. The objective is to recover the 12~parameters (5~astrometric and 7~Keplerian) describing the star position at each epoch of observation (Figure~\ref{fig:intermediate-astrometry}), making the (reasonable) assumption that other deterministic effects incorporated into AGIS (aberration, relativistic light bending, and perspective acceleration) are fully accounted for. 

For this investigation we recast the classical orbit elements ($a$, $e$, $P$, $t_{\rm p}$, $i$, $\Omega$, $\omega$) into an equivalent set consisting of the four Thiele--Innes constants ($A$, $B$, $F$, $G$), together with the frequency $f=1/P$, eccentricity $e$, and mean anomaly at the reference epoch, $M_0$. While certain aspects of this orbit fitting are rather standard, a number of specific considerations are relevant for Gaia, such as the range of the period search, and the use of a prior density of the orbit eccentricity. Further details are given in Appendix~A. 

Formal uncertainties of the fitted orbit parameters can be quite misleading, due to the strongly non-linear nature of the fitting procedure, and we use instead a Monte Carlo approach. For a given system (with fixed orbit and observation geometry), we generate $N$ observation sets with independent noise realizations, leading to $N$ different sets of estimated orbit parameters, from which their precision can be estimated. Depending on the kind of investigation, $N$ could range from 1 (when generating the statistics for a large sample of different systems), to 100 or more (when assessing the precision of a specific system).

For illustration, we show results from the Monte Carlo simulations for two transiting systems, i.e.\ for two systems with `true' $i=90^\circ$, and randomly assigned elements $\omega$, $\Omega$, and $t_{\rm transit}$. Remaining parameters were simulated as described in Section~\ref{sec:occurrence}.

\begin{figure}[ht]
\centering
\includegraphics[width=0.85\linewidth]{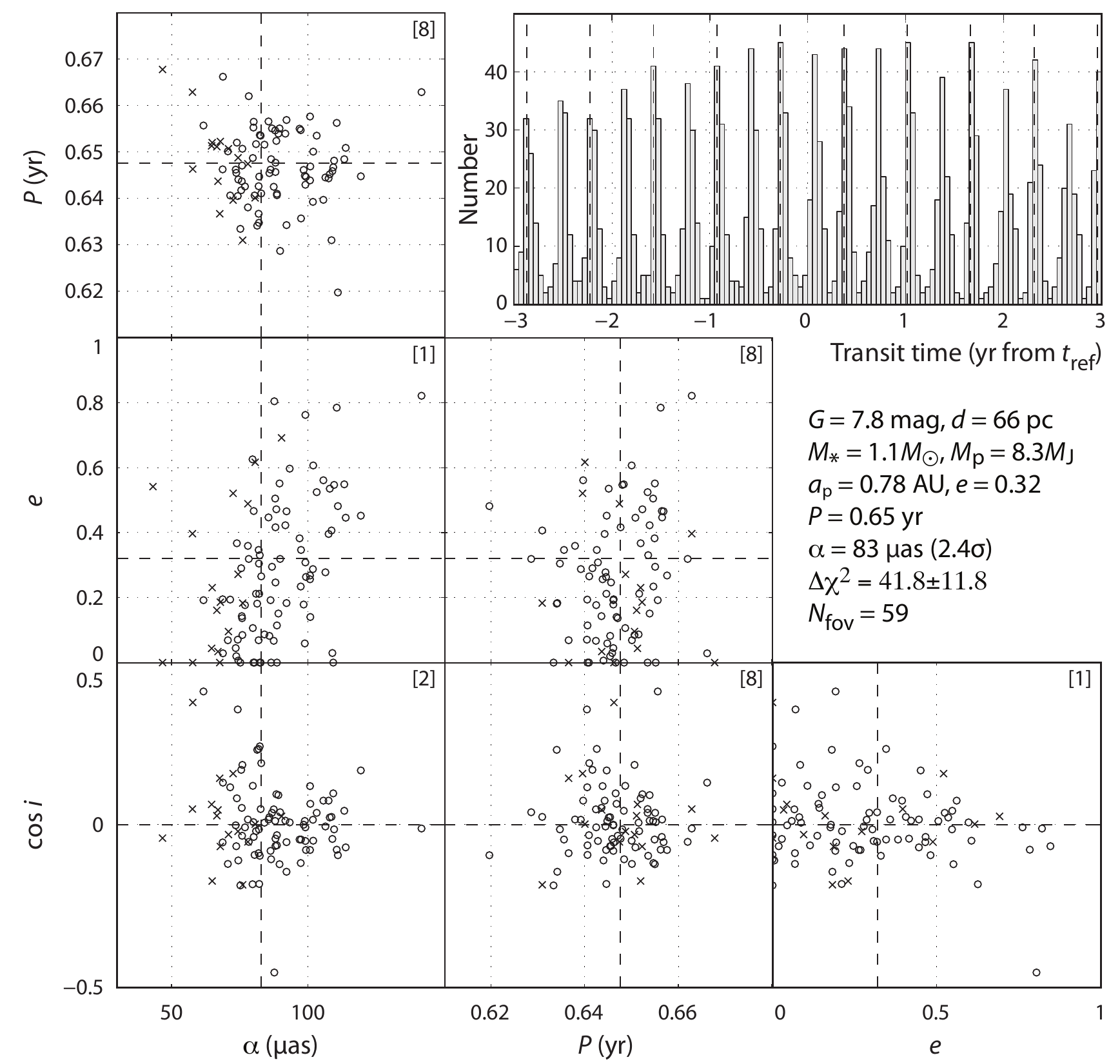}
\caption{Orbit fits to 100~simulations of one of the astrometric transiting planets (with $\Delta\chi^2=41.8\pm 11.8$), showing scatter plots of the parameters $\alpha$ (Equation~\ref{equ:astrometric-signature}), $P$, $e$, and $\cos i$, along with (top right) the transit time displacement from $t_{\rm ref}$. In all diagrams the long dashed lines show the true values. There were 17 non-detections (defined as $\Delta\chi^2<30$) among the 100 experiments. These are marked with crosses (instead of circles). With reasonable scales some of the points fall outside the plotted areas. The number (percentage) of such outliers is shown in brackets in the top right corner of each subplot.
}
\label{fig:lindegrensims-20140418-002}
\end{figure}

Our first example has $G=7.8$\,mag, $P=0.65$\,yr, $e=0.32$, and $\alpha=83$\muas, corresponding to a detection at $\alpha=2.4\,\sigma_{\rm fov}$. The number of field crossings (including 20\% dead time) is $N_{\rm fov}=59$. Figure~\ref{fig:lindegrensims-20140418-002} shows scatter plots of $a_{\rm p}$, $P$, $e$, and $\cos i$, and the predicted transit times, for 100 different noise realizations (in all diagrams the long dashed lines show the true values). The predicted transit times are for $\omega+\nu=90^\circ$ or $270^\circ$, where $\nu$ is the true anomaly. Because astrometry alone cannot determine $\omega$ unambiguously, we assume $0\le\omega<180^\circ$, and consequently have to predict two possible transit times per estimated orbit period, but only one true transit time [long-dashed line] per true period. 

\begin{figure}[ht]
\centering
\includegraphics[width=0.85\linewidth]{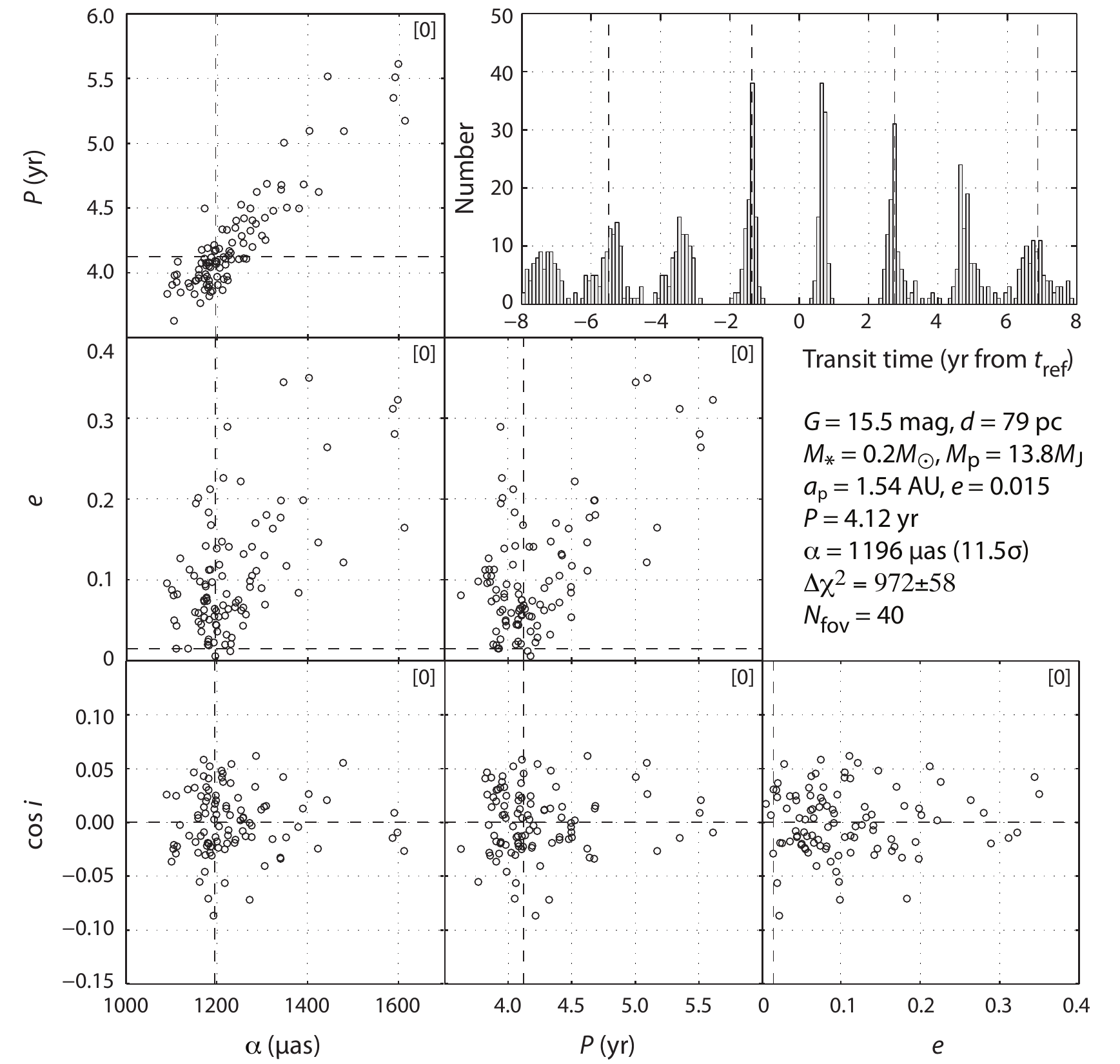}
\caption{As Figure~\ref{fig:lindegrensims-20140418-002} for another of the astrometric transiting planets, with $\Delta\chi^2=972\pm 58$. Noteworthy is the high S/N and well-behaved solution despite the relatively faint magnitude ($G=15$), the relatively long orbit period, and the small number of field crossings. In this case there were no non-detections, and no outliers. 
}
\label{fig:lindegrensims-20140418-003}
\end{figure}

Our second example (Figure~\ref{fig:lindegrensims-20140418-003}) has $G=15.5$\,mag, $P=4.12$\,yr, $e=0.015$, $\alpha=1196$\muas\ ($11.5\sigma_{\rm fov}$), and $N_{\rm fov}=40$. Despite its unusually small number of field crossings (even for its ecliptic latitude of $\beta=-5^\circ$, cf.\ Table~\ref{tab:fov-transits}) and fainter magnitude ($G=15.5$), this orbit is even better-determined due to the much higher S/N, and the larger ratio of $M_{\rm p}$ ($=13.8M_\Jupiter$) to $M_\star$ ($=0.2M_\Sun$). Interestingly, $\cos i$ is about equally well determined in all solutions, even when $a_{\rm p}$, $P$, or $e$ is significantly wrong. We also note that the predictions for $t_{\rm transit}$ are only good for a few times $P$ around the Gaia observing epoch, because of the relatively large uncertainty in~$P$. This will always be a problem for periods larger than a few years, but one that can be improved by radial velocity observations to constrain the period.

\subsection{The $\boldsymbol{\Delta\chi^2}$ metric}
\label{sec:delta-chi2}

As discussed in Section~\ref{sec:detectability-sn} the astrometric S/N ratio ($\alpha/\sigma_{\rm fov}$) is a useful zero-order indicator of detectability, but one which in reality depends on the number and distribution of observations, the inclination and eccentricity of the system, and the orbital period in relation to the total length of the observations. We will show that a more precise criterion is given by the likelihood ratio, or equivalently the reduction in the minimum $\chi^2$, when going from the 5-parameter solution to the 12-parameter solution.

Let $\chi^2_{\rm min}({\rm 12~parameter})$ be the minimum $\chi^2$ obtained when adjusting all 12~parameters (Equation~\ref{equ:chi2}). Omitting the orbit parameters (i.e.\ setting $\boldsymbol{y}=\boldsymbol{0}$ in Equation~\ref{equ:chi2}) and fitting only the astrometric parameters ($\boldsymbol{x}$) results in a fit with $\chi^2\,({\rm 5~parameter}) \ge \chi^2_{\rm min}({\rm 12~parameter})$. The increase in $\chi^2$ when omitting the orbit parameters is
\begin{equation}
\label{equ:DeltaChi2}
\Delta\chi^2 = \chi^2_{\rm min}({\rm 5~parameter}) - \chi^2_{\rm min}({\rm 12~parameter}) \, ,
\end{equation}
which can therefore be used as a test statistic for the significance of the orbit. In contrast to the S/N this quantity can be calculated without knowledge of the orbit, and is therefore applicable to real data. We note that $\Delta\chi^2$ can be small even when $\alpha/\sigma_{\rm fov}$ is very large, e.g.\ if the observations cover only a small part of the orbit. On the other hand, for a fixed S/N per field crossing, $\Delta\chi^2$ increases with the number of observations. 

Since $\exp(-\chi^2/2)$ is proportional to the likelihood of the model (assuming Gaussian observation noise with the stated standard deviation), this is effectively a likelihood ratio test and therefore close to optimal in terms of its power to detect orbital motion. Based on Wilks' theorem \citep[e.g.][]{1983kendall}, the distribution of $\Delta\chi^2$ in the absence of a companion is expected to follow the chi-squared distribution with 7~degrees of freedom. This provides a useful guide for setting the detection threshold. A possible criterion could be $\Delta\chi^2>30$, with a reasonably low probability of false detection (theoretically $\sim 10^{-4}$). Higher values of $\Delta\chi^2$ give both a more reliable detection and a higher precision of the estimated orbit. In practice, solutions with $\Delta\chi^2\simeq 30$ may be considered marginal, while those with $\Delta\chi^2>50$ are generally found to be reliable and $\Delta\chi^2>100$ typically gives orbital parameters determined to 10\% or better. The two examples shown in Figures~\ref{fig:lindegrensims-20140418-002} and \ref{fig:lindegrensims-20140418-003} have $\Delta\chi^2\simeq 42$ and 970, respectively. 

For our present primary purpose of assessing detectable numbers, we proceed by provisionally selecting systems above a certain S/N per field crossing (e.g.\ $\alpha>2\sigma_{\rm fov}$, cf.\ Table~\ref{tab:astrom-detections}). We then confirm these provisional detections by also requiring that the fit improves significantly, as measured by $\Delta\chi^2$, when proceeding from a 5-parameter to a 12-parameter solution. 

The estimation of the number of detectable systems is vastly sped up by using $\lambda+7$ as a proxy for $\Delta\chi^2$. Here, $\lambda$ is the noncentrality parameter (Appendix~B) which can be calculated, for simulated data, without actually fitting an orbit. All statistics reported below for $\Delta\chi^2>X$ were in fact derived using the criterion $\lambda+7>X$.

This additional $\Delta\chi^2$ criterion is reflected in the bottom part of Table~\ref{tab:astrom-detections}. We can see, for example, that the 16\,668 planets provisionally detected according to $\alpha>2\sigma_{\rm fov}$ result in 12\,893 secure detections according to $\Delta\chi^2>30$, or to 10\,297 detections according to the more restrictive $\Delta\chi^2>50$. This substantiates our claim (Section~\ref{sec:detectability-sn}) that $\alpha>2\sigma_{\rm fov}$ rather than $\alpha>3\sigma_{\rm fov}$ provides a reasonable zero-order estimate of the numbers detectable. In practice, final detection results are then somewhat insensitive to the actual choice of $\alpha/\sigma_{\rm fov}$, in the sense that additional provisional candidates revealed by lowering the S/N threshold are in any case subject to confirmation by the $\Delta\chi^2$ criterion. Below $\alpha\la0.5\,\sigma_{\rm fov}$ more {\it candidates\/} naturally continue to be selected, but the vast majority fail to result in additional confirmed detections.  

Our best estimates, pre-selected with $\alpha>0.5\,\sigma_{\rm fov}$ (lower thresholds are evidently required when assessing results for a 10-yr mission duration), are indicated in bold in the lower part of Table~\ref{tab:astrom-detections}. At $\Delta\chi^2>50-30$ we find 14\,806--27\,505 ($\sim21\,000\pm6000$) systems discoverable over the nominal 5-yr mission duration. As we will quantify in Section~\ref{sec:transiting}, some 25--42 of these are likely to be transiting.

\subsection{Dependency on mission lifetime}

Gaia has the potential to observe for considerably longer than the nominal 5~years (currently limited by its cold-gas attitude control mass), and a longer mission would bring very substantial improvements for exoplanet detection. Specifically: 
(a)~it will be possible to detect and reconstruct orbits with periods $P\ga5$\,yr; 
(b)~a larger number of systems with $P<5$\,yr will be detected;
(c)~the number of false detections will decrease; and 
(d)~the accuracy of the orbit solutions will be greatly improved. 
To quantify this we have repeated the orbit determinations for the 3500 simulated systems described in Section~\ref{sec:transiting} (viz., 100 realizations of 35~predicted transiting astrometric detections with S/N\,$>2$) and an assumed mission length of 10\,yr. Resulting $\Delta\chi^2$ are typically a factor 3--4 larger. 

Table~\ref{tab:astrom-detections} contains our estimated detection numbers at various levels of S/N and $\Delta\chi^2$ (to obtain complete statistics S/N ratios down to 0.5 per field crossing must be considered). We find that about 3--4 times as many systems could be detected with a comparable $\Delta\chi^2$, and that the subset of transiting systems would increase by the same factor. We find 90\,751, 53\,015, and 25\,958 detections at $\Delta\chi^2>30$, 50, and 100 respectively, with the corresponding numbers of transiting planets being 135, 82, and 31 respectively.

\section{Transiting exoplanets from Gaia astrometry}
\label{sec:transiting}

As noted in Section~\ref{sec:astrometric-signature}, few if any of the exoplanets discovered to date by ground- or space-based photometric transit searches will induce measurable displacements on their host stars (due to their typically small semi-major axes and low masses). Long-period transiting planets will also  probably remain rather elusive even with future transit surveys. Thus the Gaia {\it photometric\/} discoveries are likely to be restricted to $P\la5-10$\,d \citep{2012ApJ...753L...1D}, the planned HATPI (with its goal of imaging $\pi$\,sr of the sky with high cadence and high photometric precision; Bakos, priv.\ comm.) and the Transiting Exoplanet Survey Satellite \citep[TESS,][]{2014JAVSO..42..234R} to $P\simeq40-50$\,d, although PLATO \citep{2013arXiv1310.0696R} should extend the discovery space out to 1\,AU or so.

In this context, we draw attention to a class of transiting planet which should be discovered from Gaia {\it astrometry}: systems with large astrometric signature and which can be inferred, statistically or explicitly from their reconstructed orbit parameters, to lie edge-on to the line-of-sight. The statistical existence of such astrometric transiting planets was also noted by \citet{2014MNRAS.437..497S} in their assessment of giant planets around M~dwarfs detectable by Gaia.

We can estimate the numbers of the astrometric discoveries that will transit as follows. Considering first the case of circular orbits, the probability for a randomly-oriented planet to be favourably aligned for a transit (or secondary eclipse) is given by the {\it solid angle\/} on the sphere swept out by the planet's shadow \citep[e.g.][]{1984Icar...58..121B}
\begin{equation}
\label{equ:transits-probability}
p=\frac{R_\star}{a_{\rm p}} \simeq 0.005\left(\frac{R_\star}{R_\Sun} \right) \; \left(\frac{a_{\rm p}}{1~{\rm AU}} \right)^{-1} ,
\end{equation}
Evaluation of $i$ and $p$ for realistic cases demonstrates the well-known result that transits only occur for $i\simeq90^\circ$, while $p$ is very small. The transit probability is independent of star distance, but the corresponding photometric accuracy decreases.

The situation is subtly different for elliptical orbits: although an eccentric planet spends the majority of its {\it time\/} at distances from the star larger than its semi-major axis, the majority of its true anomaly, $\nu(t)$, is spent at smaller distances. This results in a larger fraction of the celestial sphere being intercepted by the planet's shadow, and a higher probability that the planet will transit (although the time spent in transit at these locations will be shorter). The transit probability is a function of both the true anomaly, $\nu$, and the polar angle from the orbit plane. From the expression for the star--planet distance as a function of $\nu$, and integrating over the planet's shadow for all values of $\omega$ \citep[][Equation~1--8]{2007PASP..119..986B} leads to the result that for planets on eccentric orbits
\begin{equation}
p=\left( \frac{R_\star + R_{\rm p}}{a_{\rm p}}\right) \; \left( \frac{1}{1-e^2}\right )  \ ,
\label{equ:barnes}
\end{equation}
where $e$ is the true eccentricity (not the projected eccentricity), and the small term $+R_{\rm p}$ includes the contribution from grazing transits. This reduces to Equation~\ref{equ:transits-probability} for $e=0$, and shows that planets on eccentric orbits are more likely to transit, by a factor $(1-e^2)^{-1}$, than those in circular orbits with the same semi-major axis. For a circular orbit, the geometric conditions for transits and secondary eclipses are identical, while for eccentric orbits transits may occur without a secondary eclipse, and {\it vice versa}.

To quantify the transit probabilities, we have assumed $R_{\rm p}=1R_\Jupiter$ for all $M_{\rm p}>0.3M_\Jupiter$. For planets below this limit, we draw $R_{\rm p}$ from the Kepler planet occurrence distribution of \citet{2013ApJ...766...81F}, then estimate the planet mass from its radius according to Equation~\ref{equ:mass-radius}. We then draw a uniform random number, $A_{\rm rand}$ in the interval $[0,1)$ for each planet, and consider it as transiting if $A_{\rm rand}<p$. Our predicted transiting numbers as a function of star distance and astrometric S/N are listed together with the total number of detections in Table~\ref{tab:astrom-detections}. Specifically, using the same provisional sample of 16\,668 recovered planets at $2\sigma_{\rm fov}$ (Section~\ref{sec:detectability-sn}), the mean number of transiting planets is $35\pm6$.  

\begin{figure}[ht]
\centering
\includegraphics[width=0.9\linewidth]{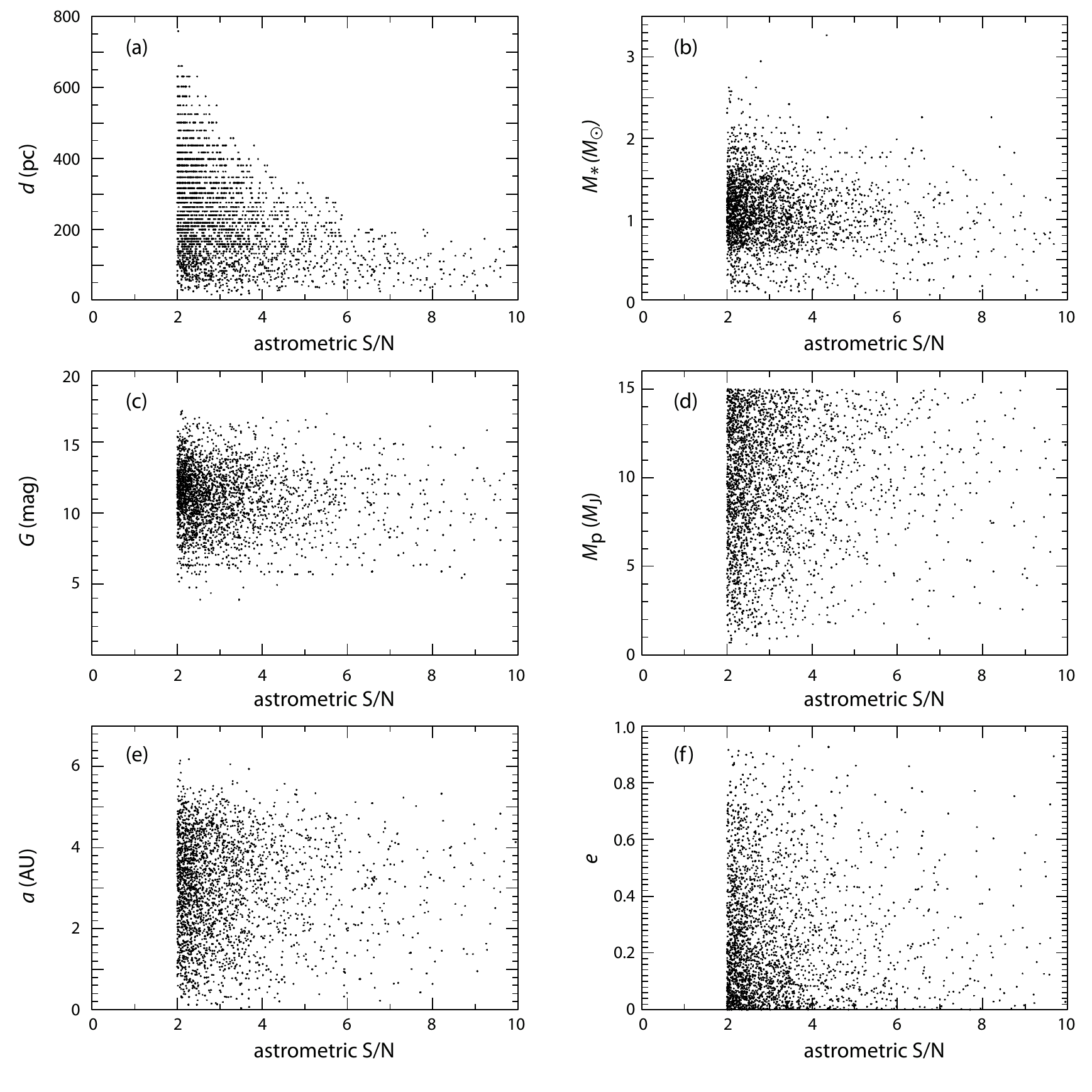}
\vspace{0pt}
\caption{The expected properties of the astrometrically-detected transiting planets as a function of astrometric S/N ($\alpha/\sigma_{\rm fov}$), based on 100 realizations of the transit samples. Even at high S/N ($\alpha/\sigma_{\rm fov}\ga8$) there is a broad distribution of expected planet properties, and we see that:
(a)~nearby planets ($d\la200$\,pc) are preferentially represented;
(b)~transiting planets will be found around low- to intermediate-mass stars ($0.5-2M_\Sun$);
(c)~detections will continue to faint magnitudes ($G\la15$).
These will span a range of:
(d)~planet masses, $M_{\rm p}\sim2-15M_\Jupiter$,
(e)~semi-major axes, $a_{\rm p}\sim1-5$\,AU, 
and 
(f)~orbit eccentricities, $e\sim0-0.8$.
}
\label{fig:scattertransiting}
\end{figure}

Figure~\ref{fig:scattertransiting} shows their expected properties as a function of the astrometric S/N ($\alpha/\sigma_{\rm fov}$) in which, to provide more representative distributions of their expected properties in view of their small numbers, we have generated 100 simulated samples of each transiting system using different random number seeds for each (thus 3500 systems in total). Being a statistical sub-sample of the detected planets, their properties represent just a corresponding scaling of those shown in Figure~\ref{fig:histogramsallplanets}. The magnitude distribution of the host stars, of relevance for the feasibility of follow-up photometric and radial velocity observations considered below, is shown in Figure~\ref{fig:scattertransiting}(c).

\begin{figure}[ht]
\centering
\includegraphics[width=0.7\linewidth]{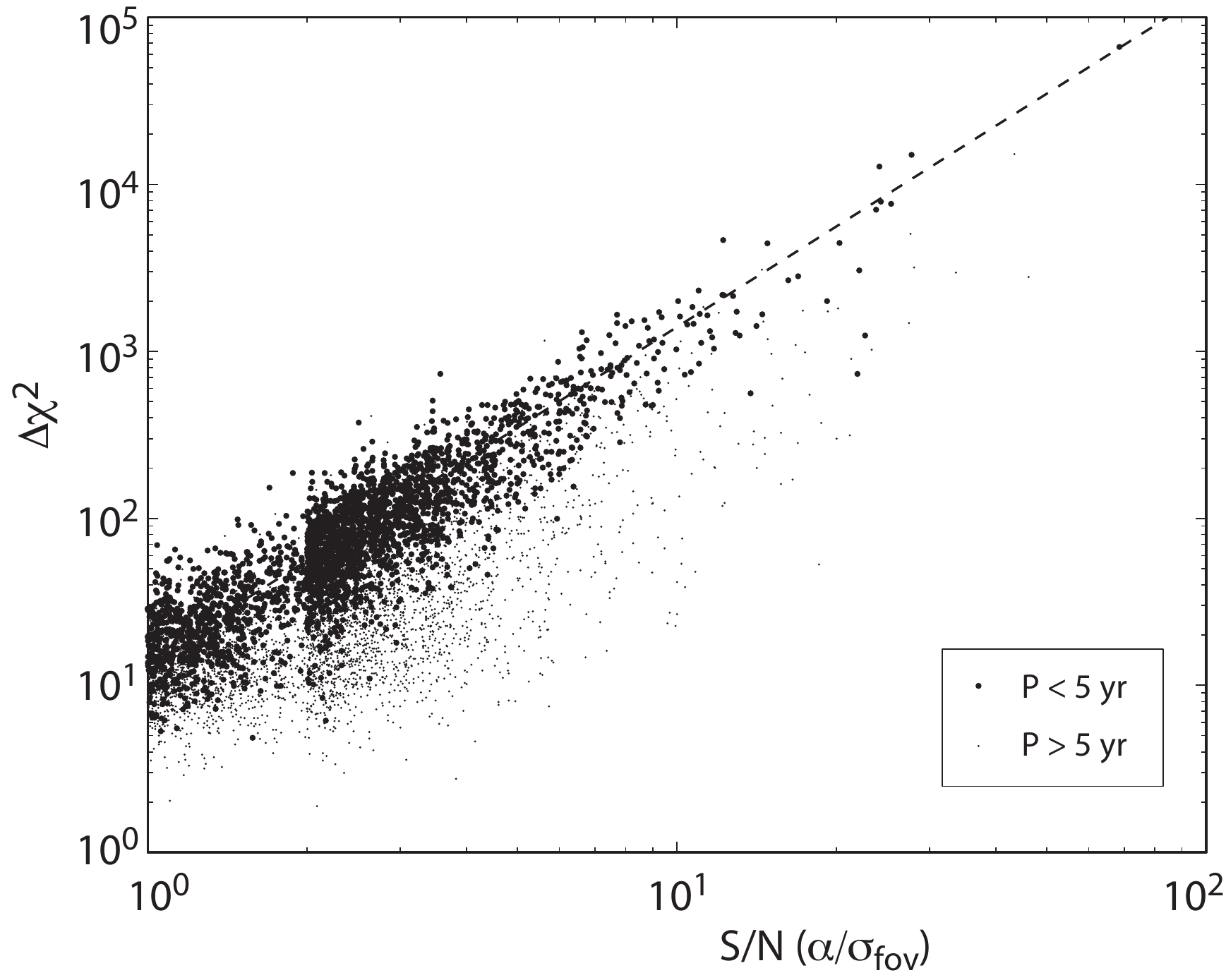}
\caption{Astrometric detectability, characterized by the quantity $\Delta\chi^2$, viz., the reduction in $\chi^2$ when going from the 5-parameter to the 12-parameter solution (Equation~\ref{equ:DeltaChi2}). In principle, $\Delta\chi^2>30$ may be considered as a reasonable detection criterion, although we use $\Delta\chi^2>100$ to identify the systems with the most accurate orbits. A mission length of 5\,yr is assumed. The trend in detectability versus S/N is rather well defined for orbit periods $P<5$\,yr, as suggested by the straight dashed line, given by $\Delta\chi^2 = 14({\rm S/N})^2$.}
\label{fig:dChi2vsSnr}
\end{figure}

Figure~\ref{fig:dChi2vsSnr} shows $\Delta\chi^2$ versus S/N for the 3500 realizations. (To emphasize the trend for small S/N, a similar number of simulations with S/N down to~1 were added; hence the discontinuity at ${\rm S/N}=2$.) For a given S/N there is a significant spread in $\Delta\chi^2$, but for systems with $P<5$\,yr the trend is much better defined, roughly corresponding to the quadratic relation $\Delta\chi^2 = 14\,({\rm S/N})^2$ shown by the straight line.

\begin{figure}[ht]
\centering
\includegraphics[width=0.7\linewidth]{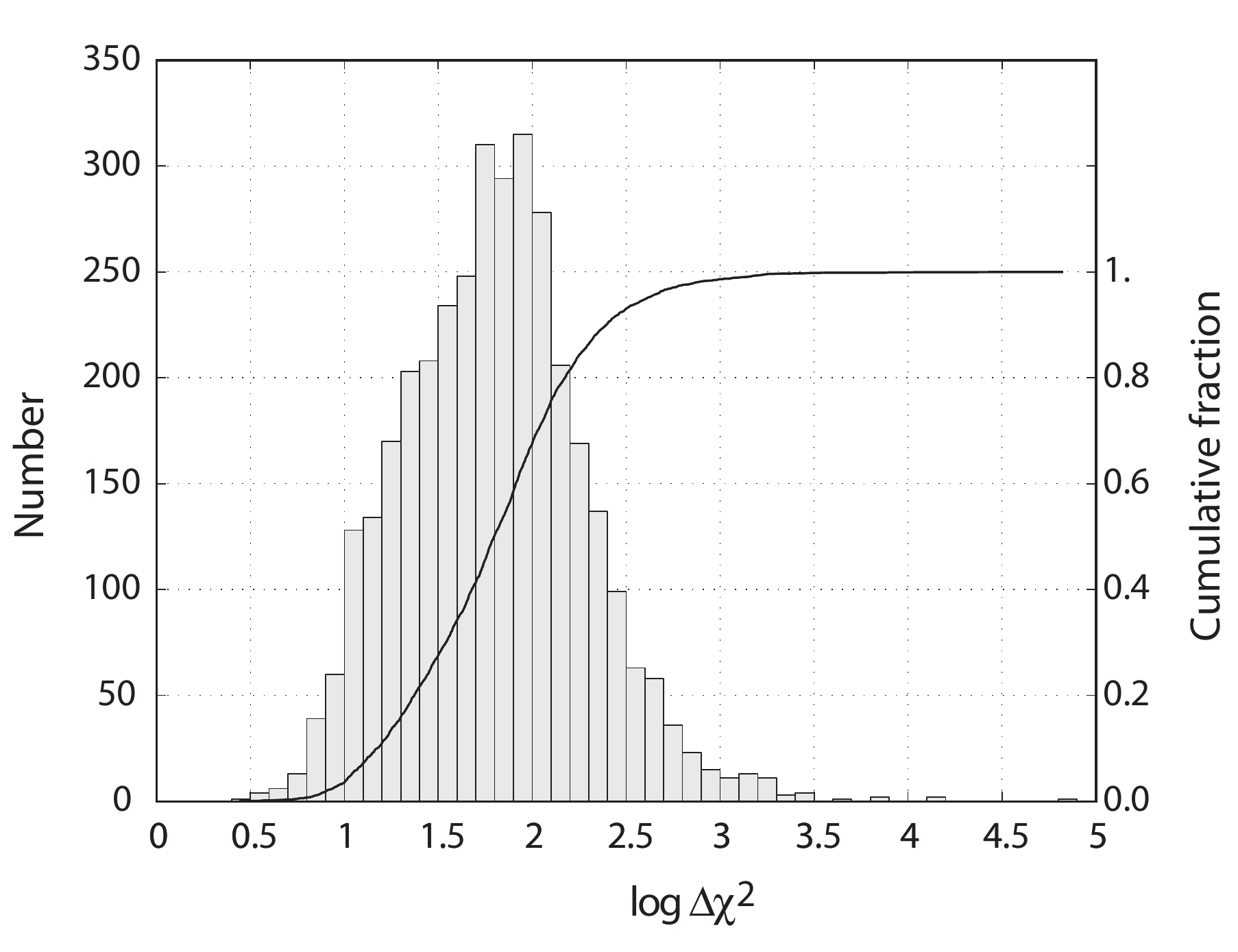}
\caption{Histogram showing the distribution of $\Delta\chi^2$ for 100 realizations of systems with transiting planets and ${\rm S/N}>2$. The solid curve is their cumulative distribution.}
\label{fig:histDChi2TP}
\end{figure}

\begin{figure}[ht]
\centering
\includegraphics[width=0.7\linewidth]{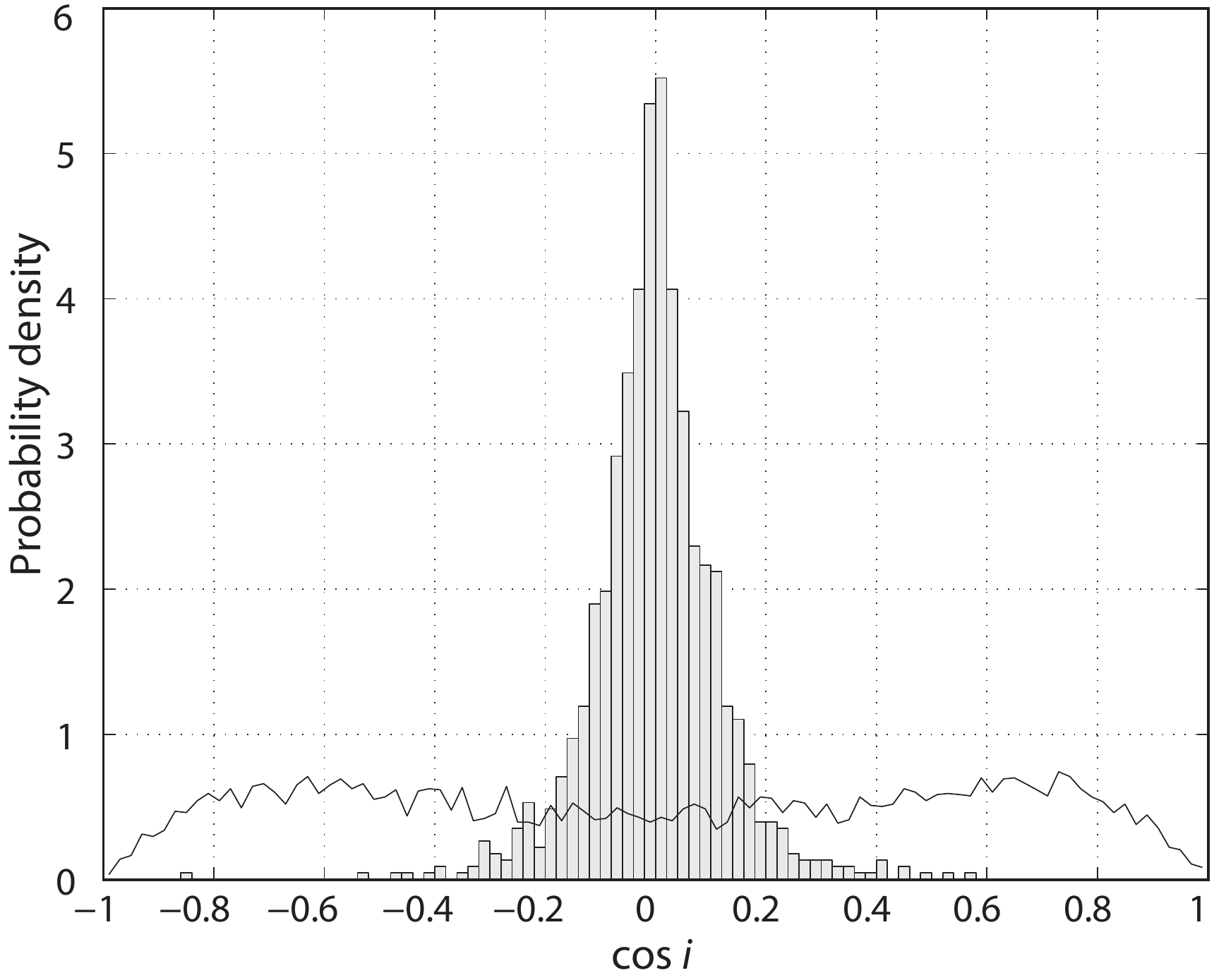}
\caption{Histogram showing the probability density of estimated $\cos i$ for transiting systems (true $\cos i\simeq 0$) with $\Delta\chi^2>100$. The curve gives the corresponding probability density function for random-inclination systems with $\Delta\chi^2> 100$.}
\label{fig:histCosi}
\end{figure}

The quality of our orbital solutions in general, and for these transiting astrometric detections in particular, improves considerably with increasing $\Delta\chi^2$. Figure~\ref{fig:histDChi2TP} shows the distribution of $\Delta\chi^2$ for the same 3500 simulations of the expected transiting systems. With the criterion $\Delta\chi^2>100$ we expect to retain 32\% of the transiting systems with S/N greater than 2, or about 11 systems. The histogram in Figure~\ref{fig:histCosi} gives the probability density of the estimated $\cos i$ for these systems. Clearly, to catch most ($\ge 80$\%) of them one must consider estimated inclinations with $|\cos i\,|\lesssim 0.1$. The time of transit is similarly estimated to within $\pm 0.1P$ for these systems, but with the ambiguity in $\omega$ this means that some 40\% of the period needs to be monitored. Depending on the period and the uncertainty in the estimated period, this fraction will increase further at epochs more removed from the mean observation epoch of Gaia. Nevertheless, the Gaia data will make it possible to identify a sample containing at least some 10~transiting systems.

The distribution of transit depths, $\Delta F\equiv (R_{\rm p}/R_\star)^2$, for the same 3500 simulated transit events is shown in Figure~\ref{fig:transitdepth-histogram}. The distribution has a median of about 0.008, increasing steeply for small values, but showing a few very pronounced transits ($\sim$1500 with $\Delta F>0.01$ and $\sim$150 with $\Delta F>0.1$, compared with the deepest currently known of $\sim$0.03 for HATS--6\,b, \citealt{2014arXiv1408.1758H}). The most prominent are long-period massive planets ($1-10M_\Jupiter$) around the nearby ($d\la100$\,pc) lowest mass ($\la0.35M_\Sun$) M~dwarfs. The systems show a mean transit duration of 0.89~day, and a mean duration as a fraction of the orbital period of 0.000\,64. 

A number of these transits may actually be present in the Gaia photometric data. To estimate the numbers, we have taken the 3500 candidate transit events (again, corresponding to 100 realizations for the 35 predicted transiting systems) and, for an improved statistical representation, made a further 10~different random initializations of the unspecified orbital parameters, subject to the constraint that it was a transiting system. We neglect the field crossing duration (about 60\,s) in comparison with the transit duration, and the effect of (possible) successive great-circle scannings during a transit event.  We found that among a total of 2\,428\,545 field crossings (corresponding to an average of 69.4 per system), 1467 ($0.042\pm0.001$ per system) occurred during transits. Thus 1~out of some 1700 field crossings of the actually transiting systems occurs during a transit (the median semi-major axis of the 3500 planet orbits is about 3\,AU, or 300$R_\star$, so that the orbit circumference is about 1800$R_\star$, roughly consistent with one transit out of 1700 random samplings). Consequently, we conclude that for our expected 25--42 astrometric detections that are predicted to transit, just one or two will have transits present in the Gaia epoch photometry itself. 

Further transits may already exist within ground-based transit data bases, and these rare but often deep transit events may also make searches for them attractive within a Planet Hunters-type human inspection of suitable photometric data sets \citep[cf.][]{2014AJ....148...28S}.  A single transit will confirm the planet and refine the orbit. 

The effort required to discover such a transit depends on the number of potential systems filtered out from the Gaia data. Assuming random inclinations (i.e.\ a uniform probability density of $\cos i$), we estimate that $\simeq 6\,500$ systems have $\Delta\chi^2>100$ (Table~\ref{tab:astrom-detections}). Of these, about 10\% or 650 will have $|\cos i\,|<0.1$ and would have to be monitored for up to 40\% of the time to find the expected $\sim 10$ transiting systems. Since we expect only 1 in every $\sim$3700 dwarfs stars to have a planet with $M_{\rm p}>0.1M_\Jupiter$ and 1\,yr\,$<P<$10\,yr, such a targeted follow-up of candidate edge-on Gaia planets would still be $\sim$50 times more efficient than a blind transit survey in terms of the number of targets that need to be observed.

There may be other prospects for improving estimates of the orbit inclination and transit times. Radial velocity observations would resolve the ambiguity in $\omega$, indicating which of the two predicted times per period should be considered. They would also provide improved estimates of $P$, improving the transit time predictions, and restricting the generally rapid divergence between predicted and true transit epochs over time which is a consequence of the rather imprecise value of $P$ derived from Gaia alone. Accurate radial velocity measures for systems showing long-period secular trends at epochs far from those of Gaia's astrometry may also assist the astrometric orbit fitting for the long-period planets. 

Properties including $e$, $M_{\rm p}$, and (inferred) densities for this unexplored class of high-mass planet orbiting at $a_{\rm p}\sim2-3$\,AU will provide useful new constraints for exoplanet formation and evolution. Additionally, they may be prime candidates for detecting planetary moons and rings \citep{2009ApJ...690....1O}, and in searching for shorter-period or slightly misaligned companions. As discussed by \citet[][Section~4.5]{2014MNRAS.437..497S} astrometric orbits in general (and for these transiting systems in particular) would also allow the epochs of favorable elongation to be identified, permitting (for the most nearby systems) optimized scheduling for imaging instruments such as Gemini--GPI \citep{2011PASP..123..692M}, VLT--SPHERE \citep{2014IAUS..299...78Z}, and E--ELT--EPICS \citep{2010SPIE.7735E..81K}.

\section{Discussion}
\label{sec:discussion}

\subsection{Dependency on spectral type}

Use of a population synthesis Galaxy model means that we have estimates of the numbers of host stars also as a function of spectral type. In this section we expand on the results for FGK dwarfs and M~dwarfs, comparing our findings to previous evaluations. 

Rather than attempting detailed predictions in view of largely unknown occurrence rates around rare spectral types more generally, we simply emphasize that the Gaia census should place many new constraints on the wider occurrence of planetary systems across the entire HR~diagram. For example, some 1.2\% of our recovered planets accompany very low mass pre-main sequence stars, while some 0.1\% (13~out of 16\,668 down to $r<17.5$ in our simulations at $2\sigma_{\rm fov}$) are around white dwarfs (identified in the $M_\star-R_\star$ plane as $R_\star\sim R_\oplus$). The numbers around white dwarfs may be conservative since we have used the current $M_\star$ (viz.\ $0.64-0.74M_\Sun$) to scale the planet occurrence rate, rather than the mass of the main sequence progenitor. More significantly, the expected numbers of white dwarfs detectable by Gaia is a strongly increasing function of the limiting magnitude: estimates suggest that some 230\,00 disk white dwarfs and some 1100 halo white dwarfs will be detected to $G=20$ \citep{1999BaltA...8..291F, 2014A&A...565A..11C}, with the majority of these within 300--400\,pc. 
We have not extended our simulations to $r>17.5$\,mag since the planet occurrence rate around white dwarfs, and especially over the relevant values of $M_{\rm p}$ and $P$, are essentially unknown. The Gaia results will provide an effective exoplanet survey for these nearby systems and will establish, for example, whether the occurrence of gas giants in wide orbits around white dwarfs is consistent with that around main sequence stars.

\subsubsection{FGK dwarfs}
\label{sec:fgkdwarfs}

\citet{2008A&A...482..699C} focused their assessement on FGK dwarfs, constrained to $V<13$ and $d<200$\,pc to provide constant astrometic precision and hence uniform Gaia detectability thresholds for their orbit-fitting experiments. With their adopted along-scan single-epoch measurement error of $\sim$11\muas\ (their $\sigma_\psi\sim8$\muas\ for successive crossings of the two fields of view) they estimated the detection of $\sim$8000 single planet systems and 1000 multiple planet systems (their Table~6). Their $3\sigma$ detection criterion would correspond to a $1\sigma$ detection for our adopted along-scan value of $\sigma_{\rm fov}=34.2$\muas\ for $G<12$. Our corresponding $1\sigma$ results around FGK stars within 200\,pc (recovered from TRILEGAL assuming 4000\,K~$<T_{\rm eff}<$~7500\,K) yields 11\,572 planet detections. 

For the same S/N-based detection criterion, these two assesments of detection numbers are therefore comparable. We stress, however, that our present overall detection numbers refer to a current Gaia astrometric performance degraded by a factor~3 with respect to that assumed by \citet{2008A&A...482..699C}. Their detection numbers for FGK stars within 200\,pc assuming the more plausible $\sigma_{\rm fov}=34.2$\muas\ (which corresponds to their $\sigma_\psi\sim24$\muas\ in their Table~6) drops to 296~single planet systems and 148~multiple planet systems.  

Notwithstanding this astrometric accuracy degradation, our {\it overall\/} planet detection numbers are significantly larger due to our inclusion of a wider range of spectral types, a fainter apparent magnitude (Figure~\ref{fig:histogramsallplanets}a) yielding more low-luminosity stars (Figure~\ref{fig:scatterallplanets}a), and our inclusion of host stars and associated planet detections beyond 200\,pc.

\subsubsection{M~dwarfs}
\label{sec:mdwarfs}

M~dwarfs provide an interesting subset of host stars for more detailed consideration, given their large numbers at relatively small distances from the Sun, their low stellar mass resulting in large astrometric displacements for a given planet mass, and their topical interest as habitable zone host stars \citep[e.g.][]{2013A&A...549A.109B}. Of the 16\,668 exoplanets recovered at $2\sigma_{\rm fov}$ for our TRILEGAL magnitude limit of $r<17.5$\,mag (Table~\ref{tab:astrom-detections}), 2097 are around M~dwarfs, identified from our simulation as stars with $M_\star<0.6M_\Sun$.

\citet{2014MNRAS.437..497S} made a detailed evaluation of the number of planets detectable around M~dwarfs with Gaia, focusing on the 33\,pc distance limit of the volume-limited LSPM sample \citep{2005AJ....130.1680L}, and on an extrapolated volume-limited sample out to 100\,pc. They adopted magnitude-dependent single measurement errors from recent Gaia models with a typical error of $\sigma_{\rm fov}=100$\muas\ for $G\la16$\,mag. They used similar criteria for astrometric detection and orbit characterization as derived by \citet{2008A&A...482..699C}. They assumed a distribution of planet occurrences defined by $M_{\rm p}=1M_\Jupiter$, $P$ uniformly distributed over $0.01-15$\,yr, and $e$ uniformly distributed over $0-0.6$, with an overall frequency of $f_{\rm p}=0.03$ \citep[as given by][]{2010PASP..122..905J}.  

As summarized in Table~\ref{tab:mdwarfs}, they predicted some 100 planets detected at $3\sigma$ within the 33\,pc horizon of the volume-limited LSPM sample (noting that LSPM is neither complete, nor entirely comprised of M~dwarfs), and 2600 detected planets (and 500 accurate orbits) out of an estimated $4\times10^5$ objects within 100\,pc based on the Besan\c con Galaxy model. They identified a statistical subset of transiting systems based on a random distribution of orbit inclinations, estimating that $\sim$40~transiting objects would be sampled out to 100\,pc, with accurate orbits for $\sim$10 (with uncertainties on $i$ of $\approx2-10^\circ$ for $i\sim90^\circ$). 

To compare these results with those from our own simulations, we proceeded as follows. Following \citet{2014MNRAS.437..497S}, we selected the 3150 reddest (in $V-J$) non-subgiant sources from the LSPM sample (not all are M~dwarfs, but \citet{2005AJ....130.1680L} does mention that some K~dwarfs are in this sample as well). Our own estimates of planet occurrences rates around M~dwarfs are dependent on both stellar mass and metallicity (Equation~\ref{equ:montet1}). For the former, we followed \citet{2014MNRAS.437..497S} in using the mass--luminosity (in $M_J$) relation. We estimated $G$ magnitudes based on the LSPM $V$~and $J$ values, adopting $V-I=0.8(V-J)-0.331$ as a fit to the M~dwarfs in \citet[][Table~2]{2011AJ....142..138L}, then using the $G(V-I)$ relation by \citet[][Table~6]{2010A&A...523A..48J}. Since metallicities are not available for all LSPM stars, we drew metallicities from the distribution of [Fe/H] values in the solar neighborhood given by \citet[][Table~3]{2001MNRAS.325.1365H}. For $R_\star$, required to determine transit probabilities, we used an analytic fit to the $M_\star-R_\star$ relation at 4.5\,Gyr, and solar metallicity, from the Dartmouth stellar models \citep{2008ApJS..178...89D}.

Finally, as detailed in Section~\ref{sec:occurrence}, we used the \cite{2014ApJ...781...28M} planet occurrence distribution for $M<0.6M_\Sun$, and the Johnson--Cumming distribution for $M>0.6M_\Sun$ (79 of our simulated stars have $M_\star>0.6M_\Sun$, with the highest mass star being $0.85M_\Sun$).  We then simulated planets as before, estimating the number of detections, and the subset of transiting systems. For the corresponding numbers for complete samples of M~dwarfs within 33\,pc and 100\,pc, we simply select stars from our TRILEGAL simulations with $M_\star<0.6M_\Sun$ within these distance limits, and run our detection and transit simulations as before.  

The various results, including those based on our $\Delta\chi^2$ detection metric, are collected in Table~\ref{tab:mdwarfs}, in the same form as for Table~\ref{tab:astrom-detections}, for two distance limits ($d<33$\,pc and $d<100$\,pc), and for two mission durations (the nominal 5\,yr, and a hypothetical 10\,yr). 

For the LSPM sample our estimate of the planets detected at $3\sigma$ is comparable to, if a little lower than, the estimates by \citet{2014MNRAS.437..497S}, viz.\ 64 and 100 respectively.

For our complete sample within 33\,pc, our number of M~stars increases by a factor~3, while the detected planet numbers increase by only 25\%: this is because most of the additional stars have very low masses (63\% have $M<0.2M_\Sun$, compared with 27\% in LSPM), while the \citet{2014ApJ...781...28M} relation identifies a decreasing planet occurrence for decreasing stellar mass. Selecting detections through the combination of a lower S/N threshold ($\alpha/\sigma_{\rm fov}>0.5$) and a $\Delta\chi^2$ exceeding 50 and 30 respectively, we estimate 121--157 detections in the complete sample out to 33\,pc, with formally only 0.2--0.3 transiting.

For the complete sample within 100\,pc, our TRILEGAL-based estimates of the total numbers of M~dwarfs ($\sim$190\,000) are lower than those of both \citet{2014MNRAS.437..497S} ($\sim$415\,000), and those implied by the northern 100\,pc M~dwarf census of \citet{2013AN....334..176L}, from which we infer a full-sky estimate of $\sim$270\,000 (including a contribution of 30\% due to incompleteness). Our corresponding numbers of planet detections at $>3\sigma_{\rm fov}$ are also somewhat less than those estimated by Sozzetti et al.\ (635~compared to~2600): while \citet{2014MNRAS.437..497S} predicted $\sim$40 transiting systems, we find essentially none ($1.1\pm1.1$). Including our $\Delta\chi^2$ metric exceeding 50 and 30 respectively, we find 1047--1451 detections out to 100\,pc, with just 1.7--2.4 transiting. Again, it can be seen from the lower part of the table that performances improve significantly for an extended 10-yr mission.

These differences in the number of M~dwarfs, and thererefore in the number of likely planets out to 100\,pc, arises in part from our magnitude limit of $r=17.5$ (compared to $G=20$ adopted by \citealt{2014MNRAS.437..497S}), such that our host star sample is incomplete for the lowest mass M~dwarfs ($<0.17M_\Sun$). Other differences originate from the Galaxy models, and the uncertainty on the known planet occurrences. Since Gaia will quantify all of these, we take this analysis no further, concluding that the \citet{2014MNRAS.437..497S} estimates are likely to be more complete, such that our estimates of the detectable planets around M~dwarfs out to 100\,pc (1047--1451) is probably underestimated, perhaps by a factor~2.

\subsection{Uncertainties on numbers}
\label{sec:uncertainties}

The uncertainties in our numbers remain rather large (with one of the goals of Gaia being to resolve these), perhaps by a factor~2 or more, due to a number of reasons.

\subsubsection{Galaxy model}

There are various uncertainties on the number and distribution of host stars given by the TRILEGAL simulations. Thus:

\noindent
(i)~the number counts are less reliable for $\vert b\vert<10^\circ$, where a smooth exponential dust disk is unrealistic, and discrete dust clouds are important. Some $172\times10^6$ (66\%) of the simulated stars to $r=17.5$, and 7500 (34\%) of the recovered planets, are at these low latitudes;

\noindent
(ii)~some 30\% more exoplanets would be detected should the star counts (and extinction) follow more closely that of the Besan{\c c}on (rather than the TRILEGAL) Galaxy model;

\noindent
(iii)~our magnitude limit of $r<17.5$\,mag means that our host star sample is incomplete for the lowest mass M~dwarfs (Section~\ref{sec:mdwarfs}), with the investigations by \citet{2014MNRAS.437..497S} suggesting that our total planet detection numbers are accordingly underestimated numerically by perhaps 1000--2000.

\subsubsection{Planet occurrence}

There are also various uncertainties on the planet distribution occurrences, including:

\noindent
(i)~the predicted yield is dominated by massive planets on wide orbits, which lie outside the range over which the distribution is well measured by radial velocity surveys, while we have assumed that the power-law holds up to $15M_\Jupiter$ and $P\sim6-10$\,yr;

\noindent
(ii)~we have not considered massive multi-planet systems, where the more complicated signals may reduce the number of planets detectable in practice;

\noindent
(iii)~our treatment of binary systems is highly simplistic. To provide estimates of planet detection numbers which are not unduly optimistic, and given the limited information on relevant planet occurrences (e.g.\ as a function of binary mass ratio, or depending on circumstellar or circumbinary orbits) and some restrictions in TRILEGAL (which does not simulated the binary star separation), we simply ignored the secondary stars of binaries, and simulated planets around the primaries according to the occurrence distributions for single stars. Even so, some 35\% of our detections are around primaries in binaries with $M_2/M_1>0.7$. Binary stars will complicate the signal, but also provide more potential host targets, with many wide binaries being resolved into separate components by Gaia. We can get an indication of the numbers of planets that may have been excluded under two extreme assumptions (and ignoring circumbinary planets): either assuming that all binaries are resolved, or that all are unresolved. In the former case, we use the $G$ magnitude of the secondary to estimate the astrometric standard error, while in the latter we use the $G$~magnitude of the combined system, and assume that $\alpha=(L_2/(L_1+L_2))\alpha_0$ where $L_i$ are the component luminosities, and $\alpha_0$ is the `undiluted' $\alpha$ for the resolved case. With these assumptions, and to be compared with the 16\,668~planets detected at $\alpha>2\sigma_{\rm fov}$ listed in Table~\ref{tab:astrom-detections}, we find 3684 recovered planets (with 2~transiting) assuming all are resolved, and 545 (with 1~transiting) if they are unresolved. At $\alpha>1\sigma_{\rm fov}$, and to be compared with the 61\,267~detections listed in Table~\ref{tab:astrom-detections}, we find 13\,201 and 2546 circumsecondary respectively; 

\noindent
(iv)~for similar reasons, for our main simulations we have simply excluded contributions from giant stars and massive main sequence stars (Section~\ref{sec:star-counts}) for which the occurrence rate of planets is poorly known empirically. We can get an indication of the numbers of planets that may have been excluded by assuming, for example, that the planet occurrence rate for non-white-dwarf stars with $\log T_{\rm eff}>4.0$ is fixed to the value at $\log T_{\rm eff}=4.0$, and by using the planet occurrence rate for giants based on the main sequence mass (but rejecting planets on orbits within the stellar envelope, $a<R_\star$). With these assumptions, and to be compared with the 16\,668~planets detected at $\alpha>2\sigma_{\rm fov}$ listed in Table~\ref{tab:astrom-detections}, we find just 412 recovered planets around giants ($\log g<3.0$), and 200 around hot stars ($\log T_{\rm eff}>4.0$), suggesting their limited effect on overall detection numbers.

\subsubsection{Instrument performance}

Instrument performances post-commissioning have been revised in mid-2014 primarily due to increased scattered light (\url{www.cosmos.esa.int/web/gaia/science-performance}). The latest assessment gives sky-averaged parallax accuracies $\sigma_\varpi=26$\muas\ and 600\muas\ for $G=15$ and 20~respectively. Compared with the pre-launch assessment (Table~\ref{tab:astrom-accuracy}) the astrometric degradation is thus negligible at $G=15$, but increases to a factor~2 at $G=20$. For $G=17$ the degradation is expected to be about 20\%. The impact on planet detection numbers estimated here should therefore be relatively small because of our adopted $r=17$ magnitude limit, and because most of our recovered systems are much brighter (cf.\ Figure~\ref{fig:histogramsallplanets}).

\subsubsection{Bright star limit}

The gating scheme for the Gaia CCDs (Section~\ref{sec:gaia-astrom-photom}) restricts integration time, and hence saturation, for stars brighter than $G\la12$. It is intended to result in a more-or-less constant measurement precision from the onset of gating to $G\sim3$ or brighter (depending on calibration techniques still to be developed). At the completion of commissioning in 2014 August, the gating scheme has been confirmed to operate nominally (J.~de Bruijne, priv.\ comm.), although it is too early for secure estimates of the precise bright star limit, or of the resulting accuracy floor in terms of calibration errors or attitude noise. 

Accordingly, we have not imposed a bright star limit in our simulations (the brightest star with a recovered planet in our particular TRILEGAL run has $G=3.9$\,mag). While access to the brightest systems will naturally be of scientific interest (for example in terms of overlap with Doppler measurements), the final exoplanet discovery {\it numbers\/} are little affected by the detailed performance at the brightest end. For example, of the 16\,668 detected planets at $\alpha/\sigma_{\rm fov}>2$ (Table~\ref{tab:astrom-detections}) just 339~are around host stars with $G<7$.

\subsubsection{Mission accuracy}

To underline the future prospects for astrometry, we note that the factor of 2.5 improved accuracy that was the original Gaia objective when accepted by the ESA advisory committees in 2000 (10\muas\ at 15\,mag) would formally extend the volume of space surveyed at a given relative distance accuracy by a factor $2.5^3$. Taking into account the scale height of the Galactic disk and the distance distribution of our detected planets we estimate that, for such a mission, all of the detection statistics presented in this paper would be scaled up by a further factor of roughly~$2.5^2\sim6$.

\section{Conclusions}

In addition to Gaia's unique determination of distances and space motions for many exoplanet systems currently known, a major additional contribution will be its unbiased magnitude-limited exoplanet census for stars of all ages, spectral types, and metallicities, with sensitivity over a range of parameter space not well studied to date.  Our re-assessment of the numbers detectable by Gaia astrometry takes into account the latest instrument performance estimates, a comprehensive host star Galaxy population model, improved estimates of exoplanet frequencies, detailed simulations of the satellite observations, and the development of a robust detection statistic based on orbit fitting. 

The $\Delta\chi^2$ test statistic that we have developed is effectively a likelihood ratio test, and is therefore close to optimal in terms of its power to detect orbital motion. It is also closely related to the orbit fitting procedures that could be applied to the real data. Novel aspects of our treatment include an eccentricity prior to impose a physically more plausible orbit (with higher predictive power in terms of orientation and phase), and the introduction of the noncentrality parameter $\lambda$ to characterize the `actual' S/N of a given orbit. For simulated data, $\lambda$ can easily be computed for millions of orbits (as it does not involve actual orbit fitting), and can thus be used to define very precisely the subset of detected systems.

Based on these considerations, we estimate that Gaia should detect, by virtue of its astrometric displacement measurements alone, some 21\,000 ($\pm6000$) planets out to $d\sim500$\,pc for the nominal 5-yr mission. At least 1000--1500 planets should be detectable around M~dwarfs out to $\sim$100\,pc. A significant fraction should have well-determined orbits, although systems with $P\ga6$\,yr (for the nominal 5\,yr mission) will be poorly constrained. 

With this large sample of astrometric discoveries, resulting insights into planet formation and evolution will include determining the properties and frequencies as a function of host star type, elucidating gas giant formation mechanisms, probing dynamical interactions and the resulting system architecture, applications to terrestrial planet studies based on the presence or absence of Jupiter-type planets, and providing much tighter constraints on population synthesis modeling in general.

An interesting statistically secure subset of the astrometric detections will be some 25--42 systems with $i\simeq90^\circ$ that should harbor transiting planets. Although identifying the transiting systems will not be straightforward due to the relatively large errors on $i$ and $t_{\rm transit}$, the resulting transit depths will often be large because of their large masses and radii, in particular for the subset of M~dwarf host stars. One or two such transits should be present in the Gaia photometry data stream. A single transit per system will provide improved prospects for estimating the radii and densities of an important class of the exoplanet population that has not been well studied to date.

Gaia has the potential to observe for considerably longer than the nominal 5~years, and we have quantified how a longer mission would bring substantial improvements for the detection, orbit determination, and period coverage. Our simulations indicate that astrometric detection and orbit characterization numbers would rise to some 70\,000 ($\pm20\,000$) exoplanets for a 10-yr mission.

\vfill
\acknowledgments

\section*{Acknowledgements}
We thank Leo Girardi for assistance with the simulations using the TRILEGAL model, Annie Robin for clarifying the bright star properties of the Besan\c con model, and Jos de Bruijne for an updated status of Gaia's bright star gating scheme. 

MP is grateful to the Department of Astrophysical Sciences, Princeton, and especially to David Spergel, Michael Strauss, and Robert Lupton, for the invitation as the 2013 Bohdan Paczy{\' n}ski Visiting Fellow during which this work was initiated, and to the K{\"a}ll\'en Committee of the Department of Astronomy and Theoretical Physics for an invitation to the University of Lund~(S), where this work was completed.
JH acknowledges partial support from NSFAST--1108686 and NASA grant NNX13AJ15G.
We thank an anonymous referee for a number of comments which have greatly improved the content and presentation.

\clearpage
\bibliography{../Michael/Exoplanet-Handbook-second-edition/biball-to-2013.bib} 



\clearpage

\appendix
\section{Orbit fitting to the Gaia astrometry}

The objective of orbit fitting is to recover the 7~Keplerian parameters, for example the classical elements $a$, $e$, $P$, $t_{\rm p}$, $i$, $\Omega$, $\omega$ (Section~\ref{sec:orbit-constraints}). For the present investigation we use an equivalent set consisting of the four Thiele--Innes constants $A$, $B$, $F$, $G$ (which are functions of $a$, $i$, $\Omega$, $\omega$; see, e.g., \citealt{1978GAM....15.....H,2011exha.book.....P}, Section~3.6) together with the frequency $f=1/P$, eccentricity $e$, and mean anomaly at the reference epoch, $M_0$. 

The use of the Thiele--Innes constants means that the least-squares problem is non-linear only with respect to the last three parameters $f$, $e$, and $M_0$. The use of $f$ instead of $P$ is convenient because the frequency resolution is largely independent of the period (other factors being constant), while the resolution in $P$ varies as $P^2$. Thus the period search, described below, can be carried out on a regular grid in frequency. The use of $M_0$ rather than $t_{\rm p}$ limits the range of this parameter to the fixed interval $[0,\,2\pi)$ and eliminates the ambiguity of $t_{\rm p}$ modulo $P$. 

The least-squares fitting of the 7~orbit parameters is done simultaneously with the fitting of the 5~parameters describing the uniform space motion of the system's center of mass: $\Delta\alpha_{*0}\equiv\Delta\alpha_0\cos\delta_0$, $\Delta\delta_0$ (the corrections to the position components at the reference epoch $t_0$), $\varpi$, $\mu_{\alpha*}$, $\mu_\delta$. This is important because some part of the orbital motion is generally absorbed by the astrometric parameters, which limits its detectability. This is particularly the case for long-period orbits ($P\gg T$), where the orbital motion may be almost completely absorbed by the proper motion parameters. Similarly, orbital motion with a period close to 1~yr may be absorbed by the parallax parameter. In tangential coordinates the linearised observation equation for the along-scan field angle $\eta$ at time $t$ is
\begin{eqnarray}
\label{equ:eta}
\Delta\eta 		&=&\left[\Delta\alpha_{*0}+(t-t_0)\mu_{\alpha*}+BX(t)+GY(t)\right]\sin\theta 				\nonumber \\
			&&\quad + \left[\Delta\delta_0+(t-t_0)\mu_\delta+AX(t)+FY(t)\right]\cos\theta + \Pi_\eta\varpi \,,
\end{eqnarray}
where $\theta$ is the position angle of the scan and $\Pi_\eta$ the along-scan parallax factor. The non-linear orbital parameters $f$, $e$, $M_0$ only appear via the functions 
\begin{equation}
\label{equ:XY}
X(t) = \cos E - e\, , \quad Y(t) = \sqrt{1-e^2} \sin E \, ,
\end{equation}
where the eccentric anomaly $E$ is obtained from Kepler's equation
\begin{equation}
E-e\sin E=M_0+2\pi f(t-t_0) \ .
\end{equation}
Given the position of an object, the AGISLab software \citep{2012A&A...543A..15H} computes a list of all expected field crossings of the object, including $t$, $\theta$, and $\Pi_\eta$ of each crossing, based on the nominal scanning law.

Equation~\ref{equ:eta} is the basis for our simulation of the observations, and for the orbit fitting. It gives, for all observations of a given object, the expected along-scan displacement of the stellar image as a function of the 12 astrometric and orbit parameters. Formally, we write this $\Delta\eta_i(\boldsymbol{x},\,\boldsymbol{y})$, where $i$ is the index of the field crossing, and $\boldsymbol{x}$, $\boldsymbol{y}$ are vectors of length 5 and 7 respectively, containing the astrometric and orbit parameters. Given the assumed set of `true' parameter values and the corresponding list of field crossings, observations are simulated as
\begin{equation}
\Delta\eta_i^{\rm obs}=\Delta\eta_i(\boldsymbol{x}^{\rm true},\,\boldsymbol{y}^{\rm true}) +\nu_i  \, , 
\end{equation}
where $\nu_i$ is Gaussian observation noise with standard deviation $\sigma_i$ (Equation~\ref{equ:sig-fov}). The simulated observations are input to the orbit fitting algorithm, in which all 12 parameters, or a subset of them, are fitted. The goodness-of-fit is measured by the chi-squared statistic 
\begin{equation}
\label{equ:chi2}
\chi^2(\boldsymbol{x},\,\boldsymbol{y}) = \sum_i \left(\frac{\Delta\eta_i^{\rm obs}-\Delta\eta_i(\boldsymbol{x},\,\boldsymbol{y})}{\sigma_i} \right)^2 \, .
\end{equation}
The least-squares solution is equivalent to the minimization of this $\chi^2$, subject to certain constraints and modifications discussed below. For Gaussian noise the likelihood of the model is proportional to $\exp(-\chi^2/2)$, so minimizing $\chi^2$ is equivalent to maximum-likelihood estimation.

For a given set of the three non-linear parameters $f$, $e$, $M_0$, the fitting of the other nine parameters ($\Delta\alpha_{*0}$, $\Delta\delta_0$, $\varpi$, $\mu_{\alpha*}$, $\mu_\delta$, $A$, $B$, $F$, $G$) is completely linear and can therefore be computed very quickly, effectively defining the goodness-of-fit as a non-linear function $\chi^2(f,\,e,\,M_0)$ of the remaining parameters. Its minimization is difficult mainly due to the many local minima obtained as a function of $f$. By contrast, for a given $f$ the dependence on $e$ and $M_0$ is much simpler, but constrained ($0\le e <1$, $0\le M_0< 2\pi$) and complicated by the degeneracy of $M_0$ for $e=0$. Mapping $(e,\,M_0)$ to the transformed pair
\begin{equation}
\label{equ:uv}
u = \frac{e}{1-e}\cos M_0\, , \quad v = \frac{e}{1-e}\sin M_0\, ,
\end{equation}
solves the degeneracy (by making $M_0$ undefined for $e=0$) and eliminates the need for a constrained minimization. Empirically, we find that for a given $f$ the topology of $\chi^2(u,v)$ is quite simple so that the minimum can be found by any standard non-linear optimization method (we used the Nelder--Mead simplex algorithm). The problem is thus reduced to finding the minimum with respect to $f$. We do this simply by searching a regular grid of frequencies $f_{\rm min} \le f \le f_{\rm max}$ in steps of $\Delta f\simeq 0.05/T$, where $T$ is the mission length. The frequency range of the search is discussed below. When the grid point with the smallest $\chi^2$ has been found, an optimum $f$ is estimated by parabolic interpolation around the minimum, and the remaining astrometric and orbital parameters are determined as described above.

The quasi-random temporal sampling produced by the Gaia scanning law (with a minimum time interval of $\sim$2\,hr between successive observations, and a minimum of 6~distinct epochs per year) in principle allows to determine orbital periods shorter than a day, but the upper limit of the frequency search, $f_{\rm max}$, must in practice be limited by the minimum physically realistic period and the risk of the global minimum not being close to the true frequency (aliasing). This risk is roughly proportional to $f_{\rm max}$ (the more frequencies that are searched, the higher is the risk of finding one that accidentally fits the data better than the true frequency), and thus inversely proportional to the shortest period in the search. It also depends strongly on the actual time sampling of the object, and is much higher in a wide band along the ecliptic (where the time sampling is relatively poor) than for objects further from the ecliptic ($|\beta|\ge 45^\circ$). In the experiments reported here we generally use $f_{\rm max}=1.6$~yr$^{-1}$, except for the relatively few ($\sim 5$\,\%) systems with true period $<0.7$\,yr, where the search is extended to 1.1 times the true frequency.

The minimum frequency searched is always $f_{\rm min}=0.016$~yr$^{-1}$ ($P\sim63$\,yr). We choose this low limit because it sometimes allows the detection of a companion from the non-linear segment of the orbit, although no orbital elements could be reliably estimated.

A common phenomenon encountered when fitting orbits with true $P>4$~yr is that the best fit is obtained for a very eccentric orbit ($e>0.7$), even though the true eccentricity is typically much smaller ($<0.3$). The minimum $\chi^2$ versus $e$ is very shallow in these cases, and the fitted high eccentricity merely an accidental effect of the noise. If a circular orbit had been adopted instead, the result would often have been a physically more plausible orbit with higher predictive power in terms of its orientation and phase. To address this problem we take a Bayesian viewpoint and maximize the posterior probability density
\begin{equation}
\label{equ:bayes}
{\rm P}(f,\,e,\,M_0) \propto {\rm P}(e)\; \exp\left[-\frac{1}{2}\, \chi^2(f,\,e,\,M_0)\right] \, .
\end{equation}  
Here ${\rm P}(e)$ is the prior density of the eccentricity, and the exponential is the (relative) likelihood of the model for Gaussian errors. Taking the logarithm of Equation~\ref{equ:bayes} we see that the prior is included by minimizing $\chi^2-2\ln\,{\rm P}(e)$ instead of $\chi^2$, resulting in a fit with a slightly larger $\chi^2$ than if the prior had not been used. All orbit fits are obtained in this way, using as prior the Beta distribution (Equation~\ref{equ:kipping-beta}) with $a=1$, $b=3$. Using $a=1$ (instead of $a=0.867$ used in Section~\ref{sec:star-counts}) regularizes the behaviour at $e=0$. For some very eccentric orbits this will bias the solution towards more circular orbits, but with the merit of avoiding a much larger number of spurious high-eccentricity solutions. For well-determined solutions the use of the prior has little influence on the fitted parameters.

\section{The noncentrality parameter, $\boldsymbol{\lambda}$}

Use of the $\Delta\chi^2$ statistic as a measure of detectability (Equation~\ref{equ:DeltaChi2}) has the advantage that it can be computed from the simulated data in the same way as for real data. For simulations it has however two disadvantages. First, it is time-consuming to compute, as it requires the fitting of orbit parameters including the period search. This makes it inconvenient in applications with large simulated samples, or for mapping the detectability over a dense grid of orbit parameters. Second, the outcome for a given system depends on the particular noise realization $\nu_i$ of the simulations, and is therefore only repeatable in a statistical sense. The latter disadvantage could be remedied by computing an {\it average\/} $\Delta\chi^2$ over many noise realizations, but with a correspondingly higher computational penalty. 

Both disadvantages can be avoided if the fitting is made to {\it noiseless\/} observations, which eliminates the random elements as well as the need for a 12-parameter fit, since the minimum $\chi^2$ in this case is~0. In the notation of Appendix~A, we may thus take
\begin{equation}
\label{equ:lambda}
\lambda = \min_{\boldsymbol{x}} \sum_i \left( 
\frac{\Delta\eta_i(\boldsymbol{x}^{\rm true},\,\boldsymbol{y}^{\rm true})-\Delta\eta_i(\boldsymbol{x},\,\boldsymbol{0})}{\sigma_i} \right)^2 
\end{equation}
as a measure of the distance between the 12- and 5-parameter models. Note that $\lambda$ is deterministic, as it does not include the observation noise ($\nu_i$). It can also be calculated very quickly as it only involves a linear least-squares fit of the astrometric parameters to the (noiseless) observations weighted by the formal standard errors. Clearly $\lambda=0$ if $\boldsymbol{y}=\boldsymbol{0}$, or more generally if the stellar motion is perfectly represented by the 5-parameter astrometric model. Naturally $\lambda$ (in contrast to $\Delta\chi^2$) can only be used with simulated data, as it requires the true parameters to be known. 

This $\lambda$ corresponds to the noncentrality parameter of the noncentral chi-squared distribution, which would be the exact distribution of $\Delta\chi^2$ for a linear model with Gaussian noise. In the linearized regime of the orbit fitting algorithm we thus expect $\Delta\chi^2$ to follow the noncentral chi-squared distribution with $k=7$ degrees of freedom and noncentrality parameter $\lambda$. This distribution has the mean value $k+\lambda$ and standard deviation $\sqrt{2(k+2\lambda)}$. 

\begin{figure}[ht]
\centering
\includegraphics[width=0.7\linewidth]{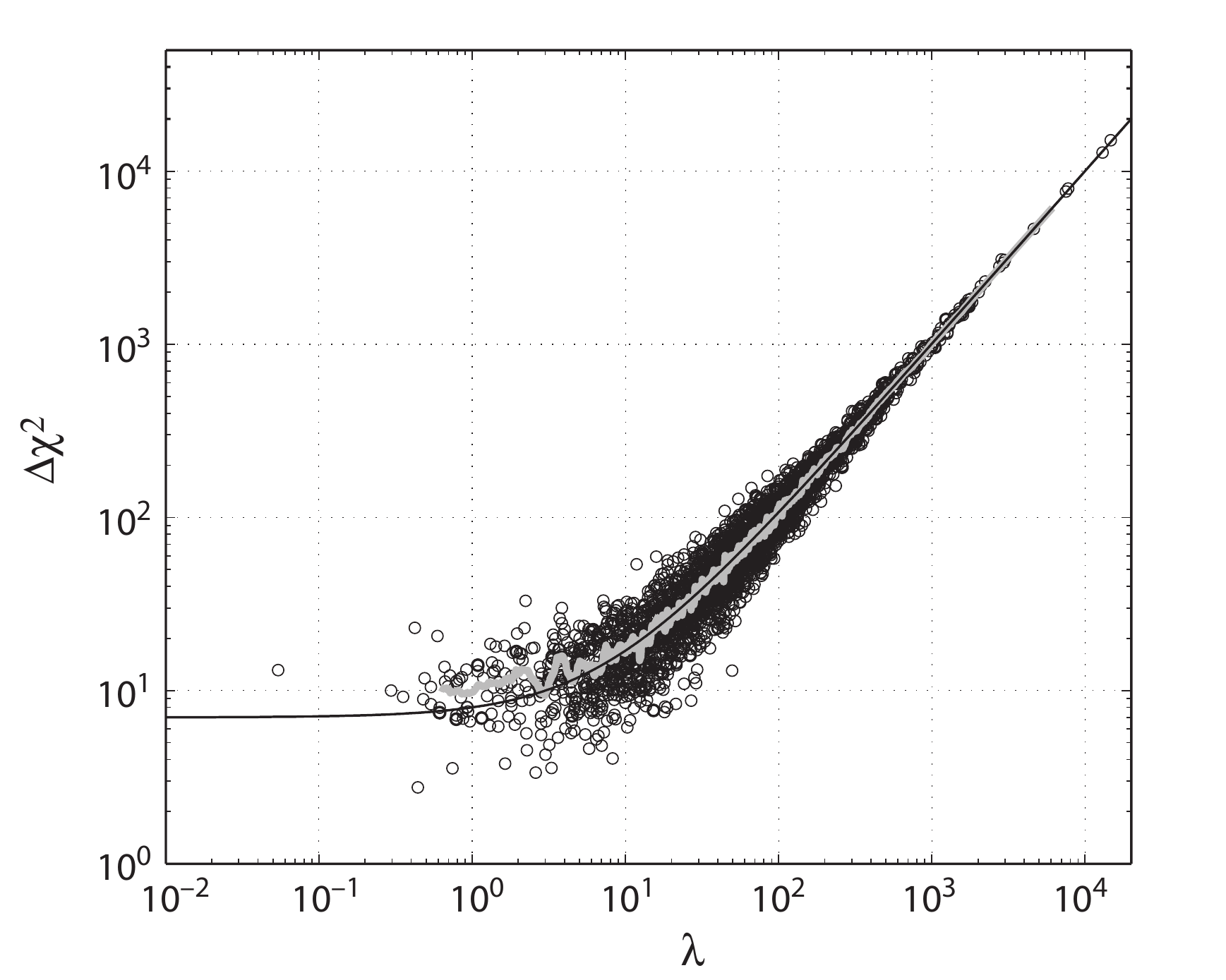}
\caption{$\Delta\chi^2$ versus the noncentrality parameter $\lambda$ for the 3500 exoplanet systems corresponding to 100~realizations of our estimated number of 35 transiting astrometric detections (circles). The thick grey curve is a 21-point running average. The thin black curve is the theoretical relation ${\rm E}(\Delta\chi^2)=\lambda+7$.}
\label{fig:dChi2}
\end{figure}

\begin{figure}[ht]
\centering
\includegraphics[width=0.7\linewidth]{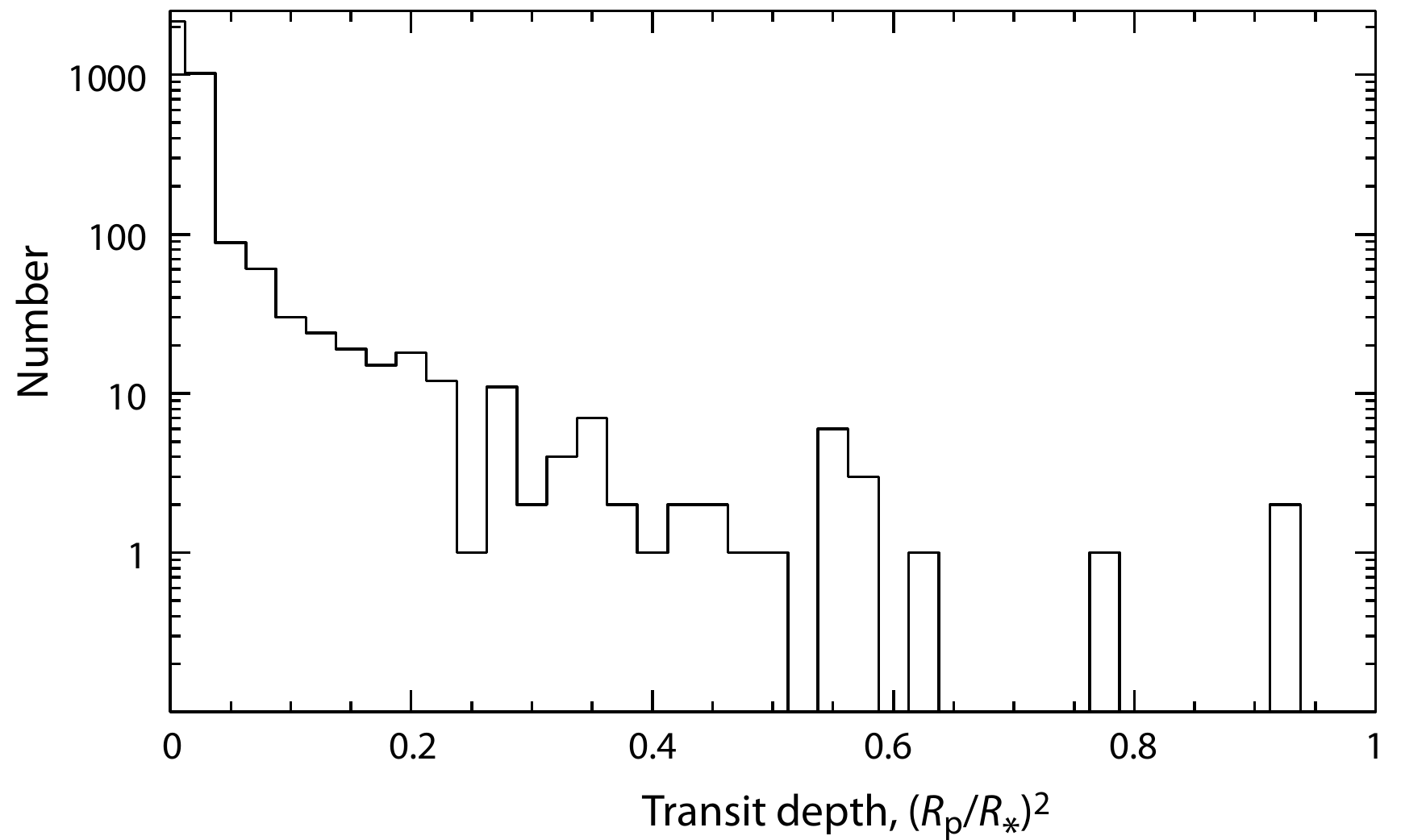}
\caption{Histogram of predicted transit depths, $(R_{\rm p}/R_\star)^2$, for the 3500 simulated transit events (corresponding to 100 realizations of the 35 transiting astrometric detections at $\alpha>2\,\sigma_{\rm fov}$ from Table~\ref{tab:astrom-detections}). The distribution has a median of about 0.008, increasing steeply for small values, but showing a few very pronounced transits attributed to long-period massive planets ($1-10M_\Jupiter$) around the lowest mass M~dwarfs. See Section~\ref{sec:transiting} for further details.}
\label{fig:transitdepth-histogram}
\end{figure}

As illustrated in Figure~\ref{fig:dChi2}, this holds to a reasonable approximation in actual orbit fitting simulations. The mean value of $\Delta\chi^2$ is slightly greater than $\lambda+7$ for small values of $\lambda$ (possibly an effect of the eccentricity prior), but agrees very well with the theoretical expectation for higher $\lambda$. The scatter of the points is also in good agreement with theory. Thus $\lambda+7$ can effectively be used as a proxy for $\Delta\chi^2$ in simulations.

\clearpage
\begin{table}[tbh]
\begin{center}
\caption{Average number of field of view passages per star, $<N_{\rm fov}>$, versus ecliptic latitude. The column $<N_{\rm fov}^\prime>$ includes a mission-level contribution of 20\% `dead time'.}			
\vspace{10pt}
\setlength\tabcolsep{15pt}
\begin{tabular}{crr}
\hline\noalign{\smallskip}
$|\,b\,|\ (^\circ)$&		$\langle N_{\rm fov} \rangle$&	$\langle N_{\rm fov}^\prime \rangle$ \\ 
\noalign{\smallskip}
\hline
\noalign{\smallskip}
0 -- 5&		64.8&		51.9 \\ 
5 -- 10&		65.4&		52.3 \\
10 -- 15&		67.2&		53.7 \\	
15 -- 20&		69.5&		55.6 \\	
20 -- 25&		73.1&		58.5 \\	
25 -- 30&		78.3&		62.6 \\	
30 -- 35&		86.4&		69.1 \\	
35 -- 40&	        99.0&		79.2 \\	
40 -- 45&        134.3&	      107.4 \\	
45 -- 50&  	      138.3&	      110.6 \\
50 -- 55&        106.3&		85.0 \\	
55 -- 60&		95.2&		76.2 \\	
60 -- 65&		88.8&		71.0 \\	
65 -- 70&		84.6&		67.7 \\	
70 -- 75&		81.5&		65.2 \\	
75 -- 80&		79.4&		63.5 \\	
80 -- 85&		78.5&		62.8 \\	
85 -- 90&		77.6&		62.1 \\	
\hline\noalign{\smallskip}
 0 -- 90&		86.1&		68.9 \\	
\noalign{\smallskip}
\hline
\end{tabular}
\label{tab:fov-transits}
\end{center}
\end{table}

\clearpage
\begin{table}[tbh]
\begin{center}
\caption{Accuracy of Gaia observations versus $G$ magnitude. 
$z$ gives the inverse relative number of photons in the image (Equation~\ref{equ:z-g}), 
$\sigma_\eta$ is the centroiding accuracy for each of the 9~along-scan CCDs in the astrometric field (Equation~\ref{equ:sig-eta}),
$\sigma_{\rm fov}$ is the along-scan accuracy per field of view passage (Equation~\ref{equ:sig-fov}),
and $\sigma_\varpi$ is the sky-averaged parallax accuracy for the nominal mission duration of 5\,yr (Equation~\ref{equ:sig-parallax}).}
\vspace{10pt}
\setlength\tabcolsep{15pt}
\begin{tabular}{rrrrr}
\hline\noalign{\smallskip}
\multicolumn{1}{c}{$G$} &	\multicolumn{1}{c}{$z$}&	\multicolumn{1}{c}{$\sigma_\eta$}&	
				\multicolumn{1}{c}{$\sigma_{\rm fov}$}&	\multicolumn{1}{c}{$\sigma_\varpi$}  \\
(mag)&	&		(\muas)&			(\muas)&			(\muas)\\
\noalign{\smallskip}
\hline
\noalign{\smallskip}
6&		0.063&	57.8&			34.2&			10.6 \\
7&		0.063&	57.8&			34.2&			10.6 \\
8&		0.063&	57.8&			34.2&			10.6 \\
9&		0.063&	57.8&			34.2&			10.6 \\
10&		0.063&	57.8&			34.2&			10.6 \\
11&		0.063&	57.8&			34.2&			10.6 \\
12&		0.063&	57.8&			34.2&			10.6 \\
13&		0.158&	91.7&			41.6&			12.9 \\
14&		0.398&	145.4&			56.1&			17.5 \\
15&		1.000&	230.9&			82.0&			25.5 \\
16&		2.512&	367.5&			125.7&			39.1 \\
17&		6.310&	588.9&			198.3&			61.7 \\
18&		15.849&	958.1&			320.6&			99.7 \\
19&		39.811&	1612.8&			538.4&			167.4 \\
20&		100.000&	2898.3&			966.5&			300.6 \\
\noalign{\smallskip}
\hline
\end{tabular}
\label{tab:astrom-accuracy}
\end{center}
\end{table}

\clearpage
\begin{table}[tbh]
\begin{center}
\caption{Summary of the adopted planet occurrence frequencies, $f$, as a function of planet mass, $M_{\rm p}$ and period, $P$. For the lower mass planets, we follow the size classification adopted by \citet{2013ApJ...766...81F}, along with the occurrence frequences given in their Table~3, and transform their adopted $R_{\rm p}/R_\Earth$ limits to $M_{\rm p}/M_\Jupiter$ using Equation~\ref{equ:fressin-mr}.
Notes: 
(a)~Fressin et al.\ give a single bin for planets in the range $6-22R_\Earth$ (in parentheses) which we interpolate to give $f=0.0114$ for the restricted range $0.1-0.3M_\Jupiter$ to avoid overlap with the region defined by the Johnson--Cumming distribution.
(b)~the frequencies apply for a Sun-like star ($1M_\Sun$, [Fe/H]\,=\,0), and have been extrapolated to cover the range $0.1-15M_\Jupiter$ and $P<10$\,yr.
}
\vspace{10pt}
\begin{tabular}{lccccl}	
\hline\noalign{\smallskip}
Class&			$R_{\rm p}$&		$M_{\rm p}$&	$		P$&			$f$&	 			Reference \\
&				($R_\Earth$)&		$(M_\Jupiter)$&		&			&				\\
\noalign{\smallskip}
\hline
\noalign{\smallskip}
Earth 		&	0.8--1.25	&			0.002--0.007&		0.8--\p85\,d& 	0.1840&		Fressin et al.\ 2013\\
super Earth 	&	1.25--2	&			0.007--0.018&		0.8--145\,d& 	0.2960&		Fressin et al.\ 2013\\
small Neptune 	&	2--4		&			0.018--0.033&		0.8--245\,d& 	0.3090&		Fressin et al.\ 2013\\
large Neptune 	&	4--6		&			0.033--0.077&		0.8--418\,d& 	0.0318&		Fressin et al.\ 2013\\
(giant		&	6--22		&			0.077--1.274&		0.8--418\,d&	(0.0524)&		Fressin et al.\ 2013$^a$)\\	
`restricted' giant&			&			0.077--0.3\p\p&		0.8--418\,d&	0.0114&		Fressin et al.\ 2013$^a$\\	
giant			&			&			0.1--0.3&			418\,d--10\,yr&	0.0388&		Johnson--Cumming$^b$\\
giant			&			&			0.3--15&			2\,d--10\,yr&	0.1339&		Johnson--Cumming$^b$\\
\noalign{\smallskip}
\hline
\end{tabular}
\label{tab:planet-frequencies}
\end{center}
\end{table}

\clearpage
\begin{table}[tbh]
\begin{center}
\baselineskip=10pt
\caption{\baselineskip=18pt
For increasing distance intervals, $N_{\rm FGK}^{\rm C08}$ lists FGK dwarf numbers from \citet[][their Table~2]{2008A&A...482..699C}, and $N_{\rm det}^{\rm C08}$ their planet detections (their Table~6). $N_\star$ gives our total star numbers from TRILEGAL (all spectral types).  $N_{\rm det}$ and $N_{\rm tran}$ are the planets detected and transiting for various S/N thresholds, $\alpha/\sigma_{\rm fov}$ (Section~\ref{sec:detectability-sn}). The lower part of the table gives the cumulative numbers, at that S/N, which {\it also\/} pass detection defined by $\Delta\chi^2$ (Section~\ref{sec:detectability-orbit}). Our best estimates of $N_{\rm det}$ and $N_{\rm tran}$ are in bold. Results are given for the nominal 5-yr mission, and for an extended 10-yr mission. }
\vspace{0pt}
\renewcommand{\arraystretch}{0.9}			
\setlength\tabcolsep{2pt}
{\footnotesize
\begin{tabular}{ccccccccccccccccccccc}		
\hline\noalign{\smallskip}
$\Delta d$&\p& $N_{\rm FGK}^{\rm C08}$&  $N_{\rm det}^{\rm C08}$&&  $N_\star$&\p&  $N_{\rm det}$&  $N_{\rm tran}$&\p& 
	$N_{\rm det}$&	$N_{\rm tran}$&\p&  $N_{\rm det}$& $N_{\rm tran}$&\p& $N_{\rm det}$& $N_{\rm tran}$&\p& $N_{\rm det}$& $N_{\rm tran}$  \\
(pc)&&   & 	&&   &&   \multicolumn{2}{c}{$\alpha>0.5\,\sigma_{\rm fov}$}&& \multicolumn{2}{c}{$\alpha>1\,\sigma_{\rm fov}$}&& \multicolumn{2}{c}{$\alpha>2\,\sigma_{\rm fov}$}&&
		 \multicolumn{2}{c}{$\alpha>3\,\sigma_{\rm fov}$}&&  \multicolumn{2}{c}{$\alpha>6\,\sigma_{\rm fov}$}  \\
\noalign{\smallskip}
\hline
\noalign{\smallskip}
\p\p0--\p50 && 	\p10\,000 & 	1400 && 	\p\p\,\p39\,000 &&						\p1508&  \p5.4&&		\p897&	\p2.8&&		\p512&	1.4&&	\p359&	0.9&&		186& 	0.4\\ 
\p50--100 && 	\p51\,000 & 	2500 && 	\p\p\,203\,000 && 						\p5914&	20.6&&		3344&	\p9.9&&		1789&	4.4&&	1195&	2.6&&		502& 	0.9\\  
100--150 && 	114\,000 & 	2600 && 	\p\p\,476\,000 && 						\p8598&	30.1&&		4877&	14.3&&		2435&	5.8&&	1466&	3.0&&		452& 	0.7\\ 
150--200 && 	295\,000 & 	2150 && 	\p\p\,889\,000 && 						11737&	37.2&&		6309&	16.7&&		2851&	6.2&&	1589&	3.1&&		289& 	0.4\\ 
200--250 && 	-- & 			-- 	&& 	\p\p\,859\,000 && 						\p8976&	26.9&&		4601&	11.5&&		1860&	3.8&&	\p862&	1.5&&		\p51& 	0.0\\ 
250--300 && 	-- & 			-- 	&& 	\p1\,298\,000 &&						11734&	33.8&&		5677&	13.5&&		2026&	4.0&&	\p832&	1.5&&		\p12& 	0.0\\ 
300--350 && 	-- & 			-- 	&& 	\p1\,793\,000 && 						14972&	42.2&&		6857&	15.9&&		2008&	3.8&&	\p636&	1.0&&		--& 		--\\ 
350--400 && 	-- & 			-- 	&& 	\p1\,775\,000 && 						13091&	35.7&&		5464&	12.4&&		1308&	2.4&&	\p264&	0.4&&		--& 		--\\ 
400--450 && 	-- & 			-- 	&& 	\p1\,411\,000 && 						\p9019&	24.3&&		3394&	\p7.5&&		\p642&	1.2&&	\p\p63&	0.1&&		--& 		--\\ 
450--500 && 	-- & 			-- 	&& 	\p1\,718\,000 && 						10439&	27.3&&		3691&	\p8.0&&		\p533&	1.0&&	\p\p25&	0.0&&		--& 		--\\ 
500--600 && 	-- & 			-- 	&& 	\p4\,267\,000 && 						21172&	53.8&&		6411&	13.9&&		\p572&	1.1&&	\p\p\p8&	0.0&&		--& 		--\\ 
600--700 && 	-- & 			-- 	&& 	\p5\,732\,000 && 						21286&	53.7&&		4984&	10.9&&		\p127&	0.3&&	--&		--&&			--& 		--\\ 
700--800 && 	-- &			-- 	&& 	\p5\,462\,000 &&						15434&	37.9&&		2678&	\p5.9&&		\p\p\p5&	0.0&&	--&		--&&			--& 		--\\ 
800--1400&&	-- &			--	&& 36\,500\,000 &&							35219&	88.0&&		2083&	\p4.7&&		--&		--&&		--&		--&&			--&		--\\
\noalign{\smallskip}
\hline
\noalign{\smallskip}
Total	&& 		470\,000&  	8750&&  62\,000\,000&&  							189099&	517&&   		61267& 	148&&		16668&  	35&&	7299&  	14&&   		1492&  	2\\
\noalign{\smallskip}
\hline
\hline
\noalign{\smallskip}
&&&&&	$\Delta\chi^2$&\p&  $N_{\rm det}$&  $N_{\rm tran}$&\p& 
	$N_{\rm det}$&	$N_{\rm tran}$&\p&  $N_{\rm det}$& $N_{\rm tran}$&\p& $N_{\rm det}$& $N_{\rm tran}$&\p& $N_{\rm det}$& $N_{\rm tran}$  \\
&&   & 	&&   &&   \multicolumn{2}{c}{$\alpha>0.5\,\sigma_{\rm fov}$}&& \multicolumn{2}{c}{$\alpha>1\,\sigma_{\rm fov}$}&& \multicolumn{2}{c}{$\alpha>2\,\sigma_{\rm fov}$}&&
		 \multicolumn{2}{c}{$\alpha>3\,\sigma_{\rm fov}$}&&  \multicolumn{2}{c}{$\alpha>6\,\sigma_{\rm fov}$}  \\
\noalign{\smallskip}
\hline
\noalign{\smallskip}
\multicolumn{5}{l}{nominal 5-year mission}&      \multicolumn{2}{c}{$>$\p30}& 		{\bf 27505}&{\bf \p42}&&		26038&	\p42&&		12893&	25&&	6541&	12&&		1475&	2\\
\multicolumn{5}{l}{\hspace{50pt}\textquotedbl}& \multicolumn{2}{c}{$>$\p50}& 		{\bf 14806}&{\bf \p25}&&		14755&	\p25&&		10297&	20&&	5762&	10&&		1444&	2\\
\multicolumn{5}{l}{\hspace{50pt}\textquotedbl}& \multicolumn{2}{c}{$>$100}& 			\p6488&	\p11&&		\p6488&	\p11&&		\p6116&	11&&	4393&	\p8&&		1353&	2\\
\noalign{\smallskip}
\multicolumn{5}{l}{extended 10-year mission}& \multicolumn{2}{c}{$>$\p30}& 		{\bf 90751}&{\bf 135}&&		58674&	117&&		16666&	35&&	7299&	14&&		1492&	2\\
\multicolumn{5}{l}{\hspace{50pt}\textquotedbl}& \multicolumn{2}{c}{$>$\p50}& 		{\bf 53015}&{\bf \p82}&&		47630&	\p80&&		16648&	35&&	7299&	14&&		1492&	2\\
\multicolumn{5}{l}{\hspace{50pt}\textquotedbl}& \multicolumn{2}{c}{$>$100}& 			25958&	\p39&&		25882&	\p39&&		15836&	30&&	7285&	14&&		1492&	2\\
\noalign{\smallskip}
\hline
\end{tabular}
}
\label{tab:astrom-detections}
\end{center}
\end{table}

\clearpage
\begin{table}[tbh]
\begin{center}
\caption{\baselineskip=18pt
Planets around M~dwarfs detectable by Gaia, for two distance limits from the Sun.  $N_\star^{\rm S14}$ and $N_{\rm det}^{\rm S14}$ are from \citet{2014MNRAS.437..497S}, for which their predicted number of astrometrically detected transiting systems are 0 (to 33\,pc) and 40 (to 100\,pc) respectively. For the 0--33\,pc volume, results are given for both the LSPM sample \citep[L; ][with $N_\star=3150$]{2005AJ....130.1680L} and for the predicted complete volume~(C), for which $N_\star$ is estimated from Galaxy models (TRILEGAL here, and Besan\c con for S14). Other details are as Table~\ref{tab:astrom-detections}. Again, our best estimates of the range of $N_{\rm det}$ and $N_{\rm tran}$ are given in bold.}
\vspace{0pt}
\renewcommand{\arraystretch}{0.9}			
\setlength\tabcolsep{2pt}
{\footnotesize
\begin{tabular}{ccccccccccccccccccccc}		
\hline\noalign{\smallskip}
$\Delta d$&\p& $N_\star^{\rm S14}$&  $N_{\rm det}^{\rm S14}$&&  $N_\star$&\p&  $N_{\rm det}$&  $N_{\rm tran}$&\p& 
	$N_{\rm det}$&	$N_{\rm tran}$&\p&  $N_{\rm det}$& $N_{\rm tran}$&\p& $N_{\rm det}$& $N_{\rm tran}$&\p& $N_{\rm det}$& $N_{\rm tran}$  \\
(pc)&&   & 	&&   &&   \multicolumn{2}{c}{$\alpha>0.5\,\sigma_{\rm fov}$}&& \multicolumn{2}{c}{$\alpha>1\,\sigma_{\rm fov}$}&& \multicolumn{2}{c}{$\alpha>2\,\sigma_{\rm fov}$}&&
		 \multicolumn{2}{c}{$\alpha>3\,\sigma_{\rm fov}$}&&  \multicolumn{2}{c}{$\alpha>6\,\sigma_{\rm fov}$}  \\
\noalign{\smallskip}
\hline
\noalign{\smallskip}
0--\p33(L) && 	\p\p3\,150& 	\p100&& 	\p\p3\,150&&						\p257&  	\p1.2&&		\p167&	0.6&&		\p93&	0.3&&		\p64&	0.2&&		\p32& 	0.1\\ 
0--\p33(C) && 	--& 			--&& 		\p\p9\,944&& 						\p344&	\p1.2&&		\p225&	0.6&&		122&		0.3&&		\p91&	0.2&&		\p42& 	0.1\\  
0--100(C) && 	415\,000& 	2600&& 	  191\,567&& 						4149&	11.1&&		2090&	4.7&&		986&		1.8&&		635&		1.1&&		267& 	0.4\\  
\noalign{\smallskip}
\hline
\hline
\noalign{\smallskip}
&&&&&	$\Delta\chi^2$&\p&  $N_{\rm det}$&  $N_{\rm tran}$&\p& 
	$N_{\rm det}$&	$N_{\rm tran}$&\p&  $N_{\rm det}$& $N_{\rm tran}$&\p& $N_{\rm det}$& $N_{\rm tran}$&\p& $N_{\rm det}$& $N_{\rm tran}$  \\
&&   & 	&&   &&   \multicolumn{2}{c}{$\alpha>0.5\,\sigma_{\rm fov}$}&& \multicolumn{2}{c}{$\alpha>1\,\sigma_{\rm fov}$}&& \multicolumn{2}{c}{$\alpha>2\,\sigma_{\rm fov}$}&&
		 \multicolumn{2}{c}{$\alpha>3\,\sigma_{\rm fov}$}&&  \multicolumn{2}{c}{$\alpha>6\,\sigma_{\rm fov}$}  \\
\noalign{\smallskip}
\hline
\noalign{\smallskip}
\multicolumn{5}{l}{0--\p33\,pc (C), 5-year mission}&	\multicolumn{2}{c}{$>$\p30}& 			{\bf \p157}& {\bf 0.3}&&	\p154&	0.3&&		114&		0.3&&	\p90&	0.2&&		\p42&	0.1\\
\multicolumn{5}{l}{\hspace{50pt}\textquotedbl}& 	\multicolumn{2}{c}{$>$\p50}& 			{\bf \p121}& {\bf 0.2}&&	\p121&	0.2&&		107&		0.2&&	\p87&	0.2&&		\p42&	0.1\\
\multicolumn{5}{l}{\hspace{50pt}\textquotedbl}& 	\multicolumn{2}{c}{$>$100}&  			\p\p87&		0.2&&	\p\p87&	0.2&&		\p85&	0.2&&	\p78&	0.2&&		\p41&	0.1\\
\noalign{\smallskip}
\multicolumn{5}{l}{0--100\,pc (C), 5-year mission}&	\multicolumn{2}{c}{$>$\p30}& 			{\bf 1451}& {\bf 2.4}&&	1415&	2.4&&		912&		1.7&&	625&		1.1&&		267&		0.4\\
\multicolumn{5}{l}{\hspace{50pt}\textquotedbl}& 	\multicolumn{2}{c}{$>$\p50}& 			{\bf 1047}& {\bf 1.7}&&	1047&	1.7&&		840&		1.5&&	595&		1.1&&		263&		0.4\\
\multicolumn{5}{l}{\hspace{50pt}\textquotedbl}& 	\multicolumn{2}{c}{$>$100}&  			\p644&		1.1&&	\p644&	1.1&&		625&		1.1&&	517&		1.0&&		255&		0.4\\
\noalign{\smallskip}
\multicolumn{5}{l}{0--\p33\,pc (C), 10-year mission}& \multicolumn{2}{c}{$>$\p30}& 			{\bf \p259}& {\bf 0.5}&&	\p219&	0.5&&		122&		0.3&&	\p91&	0.2&&		\p42&	0.1\\
\multicolumn{5}{l}{\hspace{50pt}\textquotedbl}& \multicolumn{2}{c}{$>$\p50}& 				{\bf \p193}& {\bf 0.4}&&	\p187&	0.4&&		122&		0.3&&	\p91&	0.2&&		\p42&	0.1\\
\multicolumn{5}{l}{\hspace{50pt}\textquotedbl}& \multicolumn{2}{c}{$>$100}& 				\p140&		0.3&&	\p140&	0.3&&		118&		0.3&&	\p91&	0.2&&		\p42&	0.1\\
\noalign{\smallskip}
\multicolumn{5}{l}{0--100\,pc (C), 10-year mission}& \multicolumn{2}{c}{$>$\p30}& 			{\bf 2609}& {\bf 4.3}&&	2033&	3.9&&		986&		1.8&&	635&		1.1&&		267&		0.4\\
\multicolumn{5}{l}{\hspace{50pt}\textquotedbl}& \multicolumn{2}{c}{$>$\p50}& 				{\bf 1862}&{\bf 3.1}&&	1782&	3.1&&		986&		1.8&&	635&		1.1&&		267&		0.4\\
\multicolumn{5}{l}{\hspace{50pt}\textquotedbl}& \multicolumn{2}{c}{$>$100}& 				1246&		1.9&&	1246&	1.9&&		956&		1.7&&	635&		1.1&&		267&		0.4\\
\noalign{\smallskip}
\hline
\end{tabular}
}
\label{tab:mdwarfs}
\end{center}
\end{table}

\clearpage

\clearpage
\vspace{10pt}
\noindent

\clearpage

\clearpage

\clearpage

\clearpage

\clearpage

\clearpage

\clearpage

\clearpage

\clearpage

\clearpage

\end{document}